\documentclass{article}

\usepackage{amsmath}
\usepackage{amsfonts}
\usepackage{amssymb}
\usepackage{graphicx}
\usepackage{color}

\usepackage{cite}
\usepackage{bm}

\DeclareMathOperator\arctanh{arctanh} 

\usepackage{enumitem}
\usepackage[margin=.75in]{geometry}
\usepackage{url}

\usepackage[normalem]{ulem}
\usepackage{xcolor} 

\definecolor{dartgreen}{RGB}{0, 105, 62}

\title{Introduction to correlation networks: Interdisciplinary approaches beyond thresholding}
\author{Naoki Masuda$^{1,2,*}$, Zachary M.\ Boyd$^{3}$, Diego Garlaschelli$^{4,5}$, and Peter J.\ Mucha$^{6}$\\
$^1$Department of Mathematics, State University of New York at Buffalo, USA\\
$^2$Institute for Artificial Intelligence and Data Science, State University of New York at Buffalo, USA\\
$^3$Department of Mathematics, Brigham Young University, USA\\
$^4$Lorentz Institute for Theoretical Physics, Leiden University, The Netherlands\\
$^5$IMT School of Advanced Studies, Lucca, Italy\\
$^6$Department of Mathematics, Dartmouth College, USA\\
$*$Corresponding author}
\date{}

\begin{document}
\maketitle

\begin{abstract}
Many empirical networks originate from correlational data, arising in domains as diverse as psychology, neuroscience, genomics, microbiology, finance, and climate science. Specialized algorithms and theory have been developed in different application domains for working with such networks, as well as in statistics, network science, and computer science, often with limited communication between practitioners in different fields. This leaves significant room for cross-pollination across disciplines. A central challenge is that it is not always clear how to best transform correlation matrix data into networks for the application at hand, and probably the most widespread method, i.e., thresholding on the correlation value to create either unweighted or weighted networks, suffers from multiple problems. In this article, we review various methods of constructing and analyzing correlation networks, ranging from thresholding and its improvements to weighted networks, regularization, dynamic correlation networks, threshold-free approaches, comparison with null models, and more. Finally, we propose and discuss recommended practices and a variety of key open questions currently confronting this field.
\end{abstract}


\section{Introduction}

Correlation matrices capture pairwise similarity of multiple, often temporally evolving signals, and are used to describe system interactions in various diverse disciplines of science and society, from financial economics to psychology, bioinformatics, neuroscience, and climate science, to name a few. Correlation analysis is often a first step in trying to understand complex systems data~\cite{Becker2023NatComputSci}. Existing methods for analyzing correlation matrix data are abundant. Very well established methods include principal component analysis (PCA) \cite{Jolliffe2002book} and factor analysis (FA) \cite{Anderson2003book,Mulaik2010book}, which can yield a small number of interpretable components from correlation matrices, such as a global market trend when applied to stock market data, or spatio-temporal patterns of air pressure when applied to atmospheric data. Another major method for analyzing correlation matrix data is the Markowitz’s portfolio theory in mathematical finance, which aims to minimize the variance of financial returns while keeping the expected return above a given threshold \cite{Markowitz1952JFinance,Bun2017PhysRep}. The model takes as input a correlation matrix among different assets that an investor can invest in.
In a related vein, random matrix theory (RMT)~\cite{mehta2004random,livan2018introduction,potters2020first} has been a key theoretical tool for estimating and analyzing economic and other correlation matrix data for a couple of decades~\cite{Bun2017PhysRep}. Various new methods for analyzing correlation matrix data have also been proposed. Examples include detrended cross-correlation analysis \cite{Podobnik2008PhysRevLett,QianLiuJiang2015PhysRevE,YuanFuZhang2015SciRep}, correlation dependency, defined as the difference between the partial correlation coefficient and the Pearson correlation coefficient given three nodes \cite{Kenett2010PlosOne,Kenett2012IntJBifuChaos}, determination of optimal paths between distant locations in correlation matrix data \cite{ZhouGozolchiani2015PhysRevLett}, early warning signals for anticipating abrupt changes in multidimensional dynamical systems including the case of networked systems~\cite{ChenLiuLiu2012SciRep,ChenOdea2019SciRep,Maclaren2023JRSocInterface}, and energy landscape analysis for multivariate time series data particularly employed in neuroscience~\cite{Watanabe2014NatComm,Ezaki2017PhilTransRSocA}.

\begin{figure}
\begin{center}
\includegraphics[width=16cm]{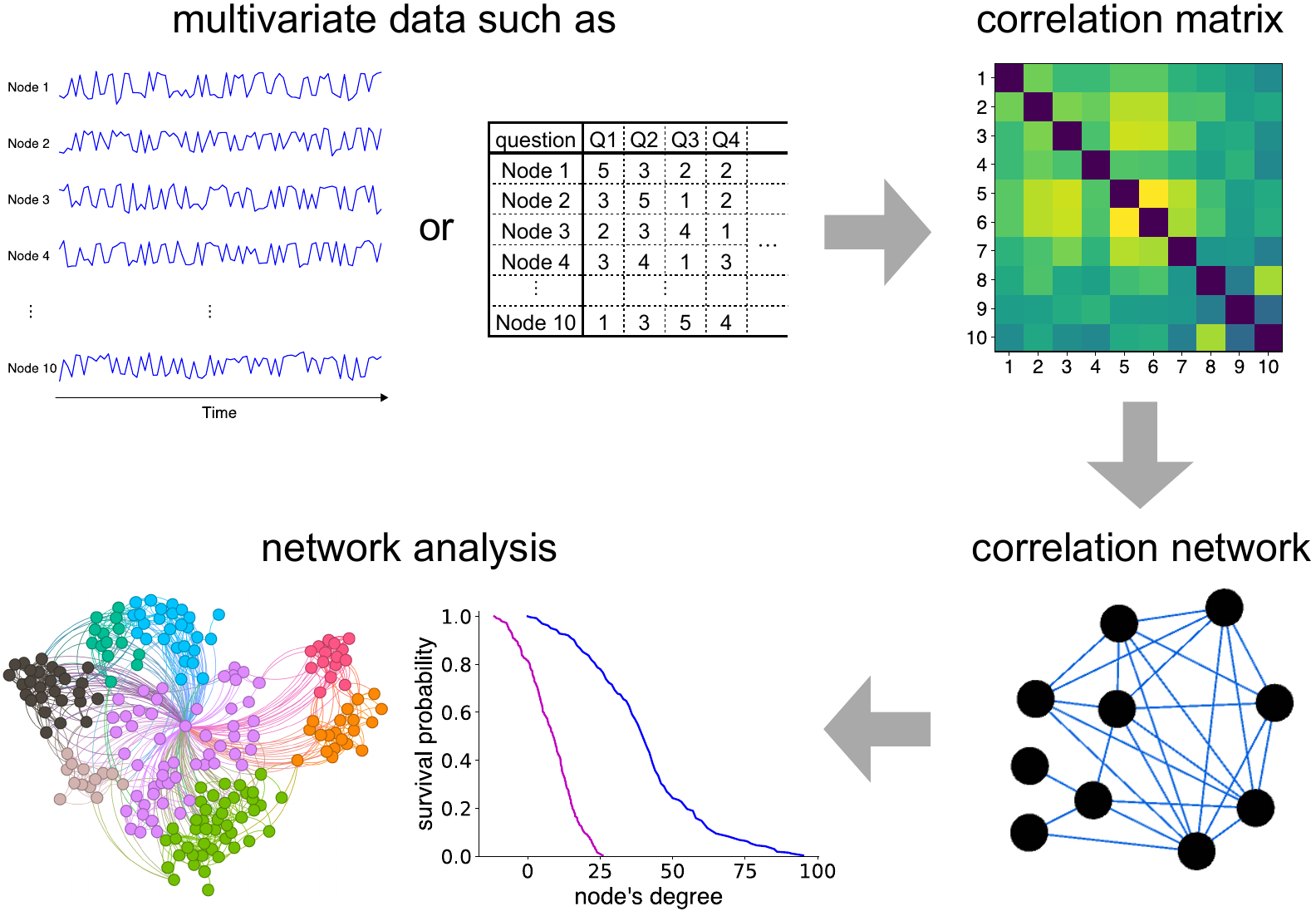}
\caption{Typical workflow of correlation network analysis. First, we are given multivariate data with $L$ samples. A sample may correspond to a time point in the case of time series data or a respondent in the case of psychological questionnaires. Second, we compute the correlation matrix. In the present article, we focus on the case of Pearson correlation matrix although the subsequent steps of analysis equally apply to the case of other types of correlation or similarity matrices. Third, we construct a correlation network from the correlation matrix by, for example, thresholding on the pairwise correlation value. Fourth, we carry out network analysis on the generated correlation network. Network analysis typically includes calculation of some quantities or network features, such as communities (also called modules) and node centrality values, for the network and assess their implications, such as the difference between a disease group of samples and healthy control group of samples. See e.g.\,\cite{Lynall2010JNeurosci, Devicofallani2014PhilTransRSocB, Borsboom2021NatRevMethodsPrimers} for similar diagrams.}
\label{fig:analysis-flow-corr-net}
\end{center}
\end{figure}

The last two decades have also seen successful applications of tools from network science and graph theory to correlational data. A correlation matrix can be mapped onto a network, which we refer to here as a \emph{correlation network}, where nodes represent elements and edges are informed by the strength of the correlation between pairs of elements. Correlation network analysis generally intends to extract useful information from data, such as the patterns of interactions among nodes or a ranking of nodes. We show a typical workflow of correlation network analysis in Fig.~\ref{fig:analysis-flow-corr-net}. With multivariate data with multiple (and not too few) samples as input, the analysis entails calculation of correlation matrices, construction of correlation networks from the correlation matrices, and downstream network analysis on the resulting correlation networks. The network analysis often includes discussion of the implication of the network analysis results in application domains in question. An ideal correlation network analysis appropriately adapts concepts and methods developed in network science to the case of correlation networks, generating knowledge that standard methods for correlation matrices (such as PCA) do not produce. Although correlation does not necessarily reflect a physical connection or direct interaction between two nodes, correlation matrices are conventionally used as a relatively inexpensive substitute of such direct connections, whose data are often less available than correlation matrix data. Correlation networks are also useful for visualization \cite{Borsboom2021NatRevMethodsPrimers}. Correlation network analysis has been used in various research disciplines, typically not much behind wherever correlation matrix analysis is used, as we will review in section~\ref{sec:appl}. In our survey here, we focus on correlation networks, with an emphasis on identifying different methods used to transform correlation matrices into correlation networks. See 
\cite{Kukreti2020FrontPhys, Tumminello2010JEconBehavOrg, Devicofallani2014PhilTransRSocB, Epskamp2018BehavRes, Borsboom2021NatRevMethodsPrimers} for shorter reviews.

The validity of correlation network analysis remains an outstanding question, especially because the decisions about how to best construct network representations from  correlation matrices is far from straightforward. One of the simplest methods is to threshold on the correlation value measured for each pair of nodes (see section~\ref{sub:dichotomizing}). However, while such a simple thresholding is widely used, it introduces various problems. These problems have led to proposals of alternative methods for generating correlation networks, which we will cover in sections~\ref{sub:weighted}--\ref{sub:graphical-lasso}. 

Before proceeding, we raise some important clarifications. First, correlation networks as we consider here are different from network architectures that exploit correlation in data~\cite{LiuLiu2022IeeeTransCircSystVideoTech,LeePark2022IeeeIcip, TuLiLiTang2022IeeeTransImageProc}\footnote{For example, the Progressive Spatio-Temporal Correlation Network (PSCC-net) is an algorithm to detect and localize manipulations in the input image data by taking advantage of spatial correlation structure in images \cite{LiuLiu2022IeeeTransCircSystVideoTech}. The superpixel group-correlation network (SGCN) \cite{LeePark2022IeeeIcip} and the deep correlation network (DCNet)~\cite{TuLiLiTang2022IeeeTransImageProc} are encoder-decoder and deep-learning network architectures, respectively, for salient object detection in images.}.
These ``networks'' are in the sense of neural network architecture in artificial intelligence and machine learning, whereas here we consider ``networks'' to denote graphs in network science.

Second, we focus on correlation networks based on the Pearson correlation coefficient or its close variants such as the partial correlation coefficient. In fact, there are numerous other definitions for quantifying the similarity between data obtained from node pairs~\cite{Vonluxburg2007StatComput, Tumminello2007IntJBifuChaos, Tumminello2010JEconBehavOrg, SmithMiller2011Neuroimage, Faust2012NatRevMicrobiol, Devicofallani2014PhilTransRSocB, WangHuang2014JTheorBiol}. Examples include similarity networks whose edges are determined using the rank correlation coefficient \cite{Fukushima2011BmcSystBiol,Millington2021PhysicaA}, the mutual information \cite{Butte2000PacificSympBiocomp,Meyer2008BmcBioinfo,GuoZhangTian2018PlosOne}, and partial mutual information
\cite{Salvador2005PhilTransRSocB,Fiedor2015ActaPhysicaPolonicaA}.
Co-occurrence of two nodes over samples, such as two authors co-authoring a paper (a paper is a sample in this example), also gives us an unnormalized variant of correlation. See sections~\ref{sub:microbiome} and \ref{sub:bibliometric} for examples of co-occurrence networks.
However, a majority of concepts and techniques explained in our main technical sections~\ref{sec:math} and~\ref{sub:null}, such as the detrimental effect of thresholding and dichotomizing the edge weight, use of weighted networks, graphical lasso, and importance of null models, also hold true when one constructs correlation networks using these or other alternative methods.

Third, we do not discuss causal inference in the present paper. A plethora of methods are available for inferring causality between nodes and associated directed networks. For example, a Bayesian network is a directed acyclic graph that fully represents the joint probability distribution of the $N$ variables. The edge of the directed acyclic graph represents directed and pairwise conditional dependency of one random variable (corresponding to a node) on another variable. The absence of the edge represents conditional independence between the two nodes. For Bayesian networks, see, e.g., \cite{Pearl2010ProcMachLearnRes, Briganti2023PsycholMethods}. For other techniques, see, e.g., \cite{WangHuang2014JTheorBiol,Runge2019NatCommun,Borsboom2021NatRevMethodsPrimers}. While these methods reveal potential causal links, even from cross-sectional data, we do not consider them further here in our discussion of correlation matrices and related general similarity matrices, whose inherently symmetric natures mean that these matrices or networks do not in principle inform us of causality or directionality between nodes (at least not in a straightforward manner \cite{Maurage2013JNutrition, Rohrer2018AdvMethodsPracticesPsycholSci, Ryan2022StructEqModeling}). In a related vein, we do not discuss time-lagged correlation of multivariate time series in this paper, since these are also asymmetric in general, although many of the same considerations we raise here also apply to lagged correlations.

\section{Data types leading to correlation networks \label{sec:appl}}

Correlation network analysis is common in many research fields. In this section, we survey typical correlation networks and their analysis in representative research fields.


\subsection{Psychological networks}

There are various multivariate psychological data, from which one can construct networks~\cite{Epskamp2018BehavRes,Borsboom2021NatRevMethodsPrimers}. For example, in personality research, researchers construct personality networks in which each node can be a personal trait or goal such as being organized, being lazy, and wanting to stay safe. Edges between a pair of nodes typically represent a type of conditional association between the two nodes.
Samples are frequently participants in the research responding to various questionnaires on a numeric scale (e.g., 5-point scale ranging from 1: strongly disagree to 5: strongly agree) corresponding to nodes. From a cross-sectional data set, one can calculate (conditional) correlation between pairs of nodes. Researchers are also increasingly combining surveys with alternative data collection modalities, for example, sensor data for daily movement or neural markers of stress \cite{Harari2016PersPsycholSci,Mcgowan2023BiolPsychiatry}. It is reasonable to use correlation between signals from different modalities (e.g., smartphones and brain scanners) to construct a correlation network \cite{Mcgowan2023BiolPsychiatry}.

Another major type of psychological network is symptom networks employed in mental health research.
Symptoms of a psychological, including psychopathological, condition, such as major depression and schizophrenia, are interrelated.
Furthermore, causality between symptoms such as fatigue, headaches, concentration problems, and insomnia, and a psychopathological disorder, is often unclear. It has been suggested that a disorder does not originate from a single root cause, which motivates the study of symptom networks~\cite{Borsboom2013AnnuRevClinPsychol,Fried2017PersPsycholSci,Fried2017SocPsychiatryPsychiatricEpid,Borsboom2021NatRevMethodsPrimers}. Nodes in a symptom network are symptoms, and one can use association between pairs of symptoms calculated from the participants' responses to define edges. Analysis of symptom networks may help us to predict how an individual develops psychopathology in the future, understand comorbidity as strong connection between symptoms of two mental disorders, and propose central nodes as possible targets of intervention \cite{Fried2017SocPsychiatryPsychiatricEpid}. Health-promoting behaviors can also be treated as nodes in these networks to suggest key behavioral intervention points~\cite{McGowan2023}.

Panel data, i.e., longitudinal measurements of variables from samples, are increasingly common for network approaches \cite{Borsboom2021NatRevMethodsPrimers}. In this case, one obtains correlation networks at multiple time points. Then, one can construct time-varying correlation networks (see section~\ref{sub:temporal}) or within-person correlation networks \cite{epskamp2018Gaussian} that reflect temporal symptom patterns and ideally expose individual differences and possible causal pathways in mental health patterns related to disorders~\cite{lutz2018network,vanzelst2022network}. 
However, the validity of psychological network approach should be further studied. Research has shown that symptom networks have poor reproducibility across samples, likely due to measurement error in assessing symptoms among other reasons~\cite{Fried2017PersPsycholSci,Forbes2017JAbnormalPsychol}.

\subsection{Brain networks\label{sec:brain}}

Various notions of brain connectivity have been essential to better understanding different neural functions. Studies of such brain networks constitute a major part of a research field that is often referred to as \emph{network neuroscience} \cite{Bullmore2009NatRevNeurosci,Bassett2017NatNeurosci}.
(See also the related material about network representations in \cite{Chung2021connectomics}.)
Multivariate time series of neuronal signals recorded from the brain are a major source of data used in network neuroscience research. Such data may be recorded in a resting state or when participants are performing some task. Functional networks or functional connectivity refer to correlation-based networks constructed from multivariate neuronal time series data, obtained through, e.g., neuroimaging or electroencephalography, where the term ``functional'' in this setting effectively means correlational. A typical node in the brain networks is either a voxel (i.e., cube of side length of, e.g., 1 mm) or a spherical region of interest (ROI), which is a sphere in the brain. In the case of multivariate time series data, there are various other methods to infer directed brain networks, which is referred to as effective connectivity, but we do not cover directed networks in this article. Brain networks based on anatomical connectivity between brain regions are referred to as structural networks. Functional connectivity, or a correlation-based edge between two nodes in the brain does not imply the presence of an edge between the same pair of nodes in the structural network. Indeed, one does not expect a one-to-one correspondence between functional and structural brain networks because several brain states and functions continuously arise from the same anatomical structure~\cite{park2013structural}. Still, the possibility of studying structural networks in combination with functional networks on the same set of nodes is a distinguishing feature of brain networks, which can be used for an additional comparison when validating the outcome of correlation-based networks~\cite{liegeois2020revisiting}. See~\cite{Bullmore2009NatRevNeurosci,SmithMiller2011Neuroimage,Craddock2013NatMethods,varoquaux2013learning,Devicofallani2014PhilTransRSocB,liegeois2020revisiting,ibanez2023noise} for reviews and comparisons of techniques for the estimation and validation of brain networks from the (partial) correlations. Examples of the use of these methods for (functional) network analysis are discussed later in this review.
%

The most typical functional neuronal networks come from neuroimaging data, in particular functional magnetic resonance imaging (fMRI) data, which are measured using blood-oxygeneration-level-dependent (BOLD) imaging techniques \cite{Biswal1995MagResonanceMed}. Functional connectivity between voxels or between spherical ROIs, or other types of nodes, is calculated by a correlation between fMRI time signals at the two nodes after one has bandpassed the fMRI time series at each node to remove artifacts, with a frequency band of, e.g., 0.01-0.1 Hz. Functional MRI improves on electroencephalogram (EEG) and magnetoencephalography (MEG) in spatial resolution at the expense of temporal resolution, but functional EEG and MEG networks are not uncommon. We also note that EEG and MEG signals are oscillatory, so one has to calculate the functional connectivity between each pair of nodes using methods that are aware of the oscillatory nature of the signal, such as using phase lag index or amplitude envelope correlation \cite{Colclough2016Neuroimage} rather than conventional correlation coefficients or mutual information.

Structural covariance networks are another type of correlation brain network where the edges are defined as the correlation/covariance of the corrected gray matter volume or cortical thickness between brain regions $i$ and $j$, where the samples are participants
%
%
\cite{Alexanderbloch2013NatRevNeurosci,Evans2013Neuroimage}. Morphometric similarity networks are a variant of structural covariance networks. In morphometric similarity networks, one uses various morphometric variables, not just a single one such as cortical thickness, for each node (i.e., ROI) \cite{Seidlitz2018Neuron}. One calculates the correlation between two nodes by regarding each morphometric variable as a sample. Therefore, differently from structural covariance networks based on cortical thickness, one can calculate a correlation network for each individual. 

In neuroreceptor similarity networks, an edge between two nodes, or ROIs, is the correlation in terms of receptor density~\cite{Hansen2022NatNeurosci}. Specifically, one first calculates a vector of neurotransmitter density for each ROI, with each entry of the vector corresponding to one type of receptor. Then, one computes the correlation between each pair of ROIs, called receptor similarity.

\subsection{Gene co-expression networks\label{sub:gene-coexpression}}

Genes do not work in isolation. Gene co-expression networks have been useful for figuring out webs of interaction among genes using network analysis methods~\cite{Horvath2008PlosComputBiol,Langfelder2008Bioinfo,Junker2008book,Vidal2011Cell,Gaiteri2014GenesBrainBehav,WangHuang2014JTheorBiol,Fiscon2018Genes,Vandam2018BriefBioinfo,mircea2022phiclust}. They are a type of data in a subfield of network science often referred to as \emph{network biology} or \emph{network medicine}. Gene co-expression networks are correlation networks in the generalized sense considered here, including the case of other measures of similarity. A typical measurement is the amount of gene expression for different genes and samples, where a sample most commonly corresponds to a human or animal  individual. If one measures the expression of various genes for the same set of samples, we can calculate the co-expression between each pair of genes by calculating the sample correlation, yielding a correlation matrix. Depending on the questions being asked in the study, it may be important to calculate the underlying correlations with different factors to account for the effects of heterogeneous gene frequencies \cite{Henderson2015,Henderson2023}. It is common to transform a correlation matrix into a network and then apply various network analysis methods, for example community detection with the aim of estimating the group of genes that are associated with the same phenotype\footnote{A \emph{phenotype} is a set of observable traits of an organism and is usually contrasted with the underlying \emph{genotype} that causes (or influences) the phenotype.} such as a disease. In this manner, correlation network analysis has been a useful tool for gene screening, which can lead to identification of biomarkers and therapeutic targets. In addition to community detection, identifying hub genes in co-expression networks helps finding key genes, for example, for cancer~\cite{Chou2014BmcGenomics}.

Different ways of defining co-expression matrices and networks from gene expression data include tissue-to-tissue co-expression (TTC) networks~\cite{Dobrin2009GenomeBiol} (also see \cite{XiangZhangHuang2013BmcGenom,Kogelman2016PlosOne}). A TTC network proposed in~\cite{Dobrin2009GenomeBiol} is a bipartite network, and its node is a gene-tissue pair. An edge between two nodes, denoted by $(\tilde{g}_i, \tilde{t}_i)$ and $(\tilde{g}_j, \tilde{t}_j)$, where $\tilde{g}_i$ and $\tilde{g}_j$ are genes, and $\tilde{t}_i$ and $\tilde{t}_j$ are tissues, represents the sample correlation as in conventional co-expression networks. However, by definition, the correlation is calculated only between node pairs belonging to different tissues, i.e., only for $\tilde{t}_i \neq \tilde{t}_j$. Therefore, TTC networks characterize co-expression of genes across different tissues.

Co-expression of genes $i$ and $j$ implies that $i$ and $j$ are both expressed at a high level in some samples (usually individuals) and both expressed at a low level in other individuals. Co-expressed genes tend to be involved in common biological functions. There are in fact multiple biophysical and non-biophysical reasons for gene co-expression~\cite{Gaiteri2014GenesBrainBehav}. For example, a transcription factor, a protein that binds to DNA, may regulate different genes $i$ and $j$ that are physically close on a chromosome. If this is the case, differential levels of regulation by the transcription factor across individuals can create co-expression of $i$ and $j$. Another mechanism of co-expression is that the expression of genes $i$ and $j$, which may be located far from each other on the chromosome or on different chromosomes, may depend on the temperature. Then, $i$ and $j$ would be co-expressed if different individuals are sampled from living environments with different temperatures.
Variation in ages of the individuals can similarly create co-expression among age-related genes.
Alternatively, co-expression may originate from non-biological sources, such as technical or laboratory ones, whose exact origins are often unknown.

One is often interested in looking for differential co-expression, which refers to the different levels of gene co-expression between two phenotypically different sets of samples, such as a disease set versus a control set, or in two types of tissues \cite{Gaiteri2014GenesBrainBehav,Vandam2018BriefBioinfo}. Differential co-expression often reveals information that one cannot obtain by examining differential expression (as opposed to co-expression), i.e., different levels of gene expression between the two sets of samples \cite{Hudson2009PlosComputBiol}.

\subsection{Metabolite networks\label{sub:metabolite}}

Metabolites are small molecules (e.g., amino acids, lipids, vitamins) that are intermediates or end products of metabolic reactions. One can also construct correlation networks from metabolomics data, or data of metabolites and their reactions \cite{Steuer2006BriefBioinfo,Silverman2020WiresSystBiolMed}.To inform the edge, one measures pairwise correlation between the amounts of two metabolites given the samples. Like gene co-expression, correlation between metabolites can occur for multiple reasons, including knock-out of a gene coding an enzyme that is involved in a chemical reaction consuming or producing two metabolites, different temperatures or other environmental conditions under which different samples are obtained, or intrinsic variability owing to cellular metabolism~\cite{Steuer2006BriefBioinfo}. Note that mass conservation within a moiety-conserved cycle produces negative correlation between at least one pair of metabolites involved in the reaction~\cite{Camacho2005Metabolomics}. That said, in some cases one may consider correlation or other similarity between only a subset of metabolites that are not necessarily associated to one another by direct chemical processes but instead draw from a set of alternative biochemical processes (see, e.g., \cite{Henderson2019}).

\subsection{Microbiome networks\label{sub:microbiome}}

Microbes interact with other microbe species as well as with their environments. Understanding of microbial composition and interaction in the human gut is expected to inform multiple diseases. Similarly, understanding soil microbial communities may contribute to enhancing plant productivity. Network analysis is adept at revealing, e.g., ecological community assembly and keystone taxa, and has been increasingly contributing to these fields.

In microbiome network analysis, one collects samples from, e.g., soil,  at various time points or locations. Each sample from an environment (e.g., soil, gut, animal corpus, or water) contains various microorganisms with different quantities. Co-occurrence network analysis is increasingly common in this field, aided by an increasing amount and accuracy of data~\cite{Faust2012NatRevMicrobiol,Hirano2019BmcBioinfo,Goberna2022SoilBiolBiochem}. In a microbiome co-occurrence network, nodes are microorganisms (e.g., bacteria, archaea, or viruses), specified at the taxa level, for example, and an edge is defined to exist if two nodes co-occur across the samples. Therefore, microbiome co-occurrence networks are essentially microbiome correlation networks, and the usual correlation measures, such as Pearson correlation, can be used to determine edge data, but more sophisticated methods to define edges are more often used. (See \cite{Faust2012NatRevMicrobiol,Hirano2019BmcBioinfo} for various co-occurrence network construction methods.) Positively weighted edges result because of, e.g., cooperation between two taxa, sharing of niche requirements, or co-colonization. Negatively edges result because of, e.g., competition for space or resources, prey-predator relationships, or niche partitioning. A historically famous example of negative co-occurrence in ecological community assembly study is the checkerboard-like presence-absence patterns of two bird species inhabiting an island, discussed by Jared Diamond \cite{Diamond1975chapter}. (Also see \cite{Faust2012NatRevMicrobiol} for a historical account.) Regardless, one should keep in mind that correlation, or co-occurrence, does not immediately imply physical interaction between two taxa.

\subsection{Disease networks}

A node in a disease network is a disease phenotype. Correlation between two diseases defines an edge, and there are various definitions of edges as we introduce in this section. Each definition of edge creates a different type of disease network.

Comorbidity, also called multimorbidity~\cite{HuThomas2016NatRevGenet}, is the simultaneous occurrence of multiple diseases within an individual. One cause of comorbidity is that the same gene or disease-associated protein can trigger multiple diseases. Other causes, such as environmental factors or behaviors, such as smoking, can also result in comorbidity. A collection of potentially comorbid diseases can be modeled as the nodes of a network, and the edges, which are based on comorbidity or other similarity index between diseases~\cite{Silverman2020WiresSystBiolMed}, are correlational in nature. 

The authors of~\cite{Hidalgo2009PlosComputBiol} constructed phenotypic disease networks (PDNs) in which nodes are disease phenotypes. The edges are sample correlation coefficients or a variant, and the samples are patients in a hospital claim record (i.e., Medicare claims in the US). Note that here one uses correlation for binary variables because each sample (i.e., patient) is either affected or not affected by any disease $i$. The authors found that, for example, patients tend to develop illness along the edges of the PDN~\cite{Hidalgo2009PlosComputBiol}.

Similarly, prior work constructed a human disease network when two diseases share at least one associated gene, which is similar in principle to the phenotypic disease network despite that the edge of the human disease network is not a conventional correlation coefficient~\cite{GohCusick2007PNAS} (also see \cite{Rzhetsky2007PNAS}). Similarly, an edge in a metabolic disease network is defined to exist when two diseases are either associated with the same metabolic reaction or their metabolic reactions are adjacent to each other in the sense that they share a compound that is not too common~\cite{LeePark2008Pnas}. (H${}_2$O and ATP, for example, are excluded because they are too common.) Alternatively, in a human symptom disease network~\cite{ZhouMenche2014NatCommun}, the edge between a pair of diseases is a correlation measure in which each sample is a symptom. In other words, roughly speaking, the edge weight is large when two diseases share many symptoms. 

\subsection{Financial correlation networks\label{sub:financial}}

Stocks of different companies are interrelated, and the prices of some of them tend to change similarly over time. A common transformation of such financial time series before constructing correlation matrices and networks is into the time series of logarithmic return, i.e., the successive differences of the logarithm of the price, given by
\begin{equation}
x_i (t) = \ln \frac{z_i(t+1)}{z_i(t)},
\label{eq:log-return}
\end{equation}
where $z_i(t)$ is the price of the $i$th financial asset at time $t$, such as the closure price of the $i$th stock on day $t$, and $t\in \{0, \ldots, T-1\}$. An advantage of this method is that $x_i(t)$ is not susceptible to changes in the scale of $z_i(t)$ over time \cite{Mantegna1999book}. Then, one constructs the correlation matrix for $N$ time series $\{x_i(1), \ldots, x_i(T-1)\}$, where $i\in \{1, \ldots, N\}$.

Financial correlation matrices have been analyzed for decades. For example, Markowitz's portfolio theory provides an optimal investment strategy as vector $\bm w = (w_1, \ldots, w_N)^{\top}$, where $w_i$ represents the fraction of investment in the $i$th financial asset, and $^{\top}$ represents the transposition~\cite{Markowitz1952JFinance,Bun2017PhysRep}. The theory formulates the minimizer of the variance of the return, $\bm{w}^{\top} C \bm{w}$, where $C$ is the covariance matrix, as the solution of a quadratic optimization problem with the constraint that the expected return, $\bm{w}^{\top} \bm{g}$, where $\bm{g} = (g_1, \ldots, g_N)^{\top}$, and $g_i$ is the expected return for the $i$th asset, is larger than a prescribed threshold.

Financial correlation matrices have also been extensively studied in econophysics research since the 1990s, with successful uses of RMT~\cite{Mantegna1999book,Laloux1999PhysRevLett,Plerou1999PhysRevLett,Bun2017PhysRep,Macmahon2015PhysRevX,almog2015mesoscopic,anagnostou2021uncovering,zema2021mesoscopic} and network methods such as maximum spanning trees \cite{Mantegna1999EurPhysJB,Bonanno2003PhysRevE}, community detection~\cite{fenn2012dynamical,almog2015mesoscopic,Bazzi2016MultModelSimul,anagnostou2021uncovering,zema2021mesoscopic}, and more advanced methods~(see \cite{Tumminello2010JEconBehavOrg, Kukreti2020FrontPhys} for reviews). One usually employs RMT in this context to verify that most eigenvalues of the empirical financial correlation matrices lie in the bulk part of the distribution of eigenvalues for random matrices. Such results imply that most eigenvalues of the empirical correlation matrices can be regarded as noise, and one is primarily interested in other dominant eigenvalues of the empirical correlation matrices whose values are not explained by random matrix theory~\cite{Laloux1999PhysRevLett,Plerou1999PhysRevLett,Macmahon2015PhysRevX,almog2015mesoscopic,anagnostou2021uncovering,zema2021mesoscopic}. The largest eigenvalue is usually not explained by RMT and is often called the market mode because it represents the movement of the market as a whole; moreover, other deviating eigenvalues are also present, encoding for the presence of groups of stocks that move coherently, as we discuss in section~\ref{sec:RMT}.

Other types of financial data are possible. For example, correlation networks were constructed from pairwise correlation between the daily time series of the investor's behavior (e.g., the net volume of Nokia stock traded or its normalized version) for two investors \cite{Tumminello2012NewJPhys,Ranganathan2018PlosOne}. One can also renormalize the covariance matrix using other indices, such as momentum~\cite{kercher}.\footnote{Momentum in finance generally refers to the rate of change in price. If the prices of two assets are correlated, the momentum of one asset can be informative of the future price of the correlated asset~\cite{momentum1}. In~\cite{kercher}, momentum and price correlation are mixed in various ways to construct correlation-type networks that reflect collective price dynamics, and, for example, network centrality is predictive of large upcoming swings in asset prices.} 

\subsection{Bibliometric networks\label{sub:bibliometric}}

Apart from microbiome studies, bibliometric and scientometric studies are another research field in which co-occurrence networks are often used~\cite{YanDing2012JAmerSocInfoSciTech, QiuDongYu2014Scientometrics}. For example, in an academic co-authorship network, a node represents an author, and an edge represents co-occurrence (i.e., collaboration) of two authors in at least one paper. One can weigh the edge according to the number of coauthored papers or its normalized variant \cite{Newman2001PhysRevE-collabo2}. While keeping authors as nodes, one can also create other types of co-occurrence networks, such as co-citation networks in which an edge connects two authors whose papers are cited together by a later paper, and keyword-based co-occurrence networks in which an edge connects two authors sharing keywords associated with their papers. Nodes of co-occurrence bibliometric networks can also be journals, institutions, research areas, and so forth. These co-occurrence networks are mathematically close to correlation networks and have been useful for understanding research communities and specialities, communication among researchers, interdisciplinarities, and the structure and evolution of science, for example.

Various other web-based information has also been analyzed as co-occurrence networks. For example, tags annotating posts in social bookmarking systems can be used as nodes of co-occurrence networks~\cite{Cattuto2009PNAS}. Two tags are defined to form an edge if both tags appear on the same post at least once. One can also use the number of the posts in which the two tags co-appear as the edge weight. Another example is co-purchase networks in online marketplaces, in which a node represents an item, and an edge represents that customers frequently purchase the two items together~\cite{LuoWangPromislow2006Wi}.

\subsection{Climate networks}

Climate can be analyzed as a network of interconnected dynamical systems \cite{Tsonis2004PhysicaA,Tsonis2006BullAmerMeteorolSoc,Donges2009Epl,FanMeng2021PhysRep}. In most analyses, the nodes of the network are equal-angle latitude-longitude grid points on the globe. However, such angular partitions lead to grid cells with geometric areas that vary with latitude, which in particular might lead to spurious correlations in the measured quantities, especially near the poles; such biases might be addressed either by a node splitting scheme that aims to obtain consistent weights for the network parameters \cite{Heitzig2012}, or by choosing instead to work on a grid with (possibly only approximately) equal grid cell areas \cite{Scarsoglio2013}. Each node has, for example, a time series measurement of the pressure level, which represents wind circulation of the atmosphere. The edge between a pair of nodes is based on the correlation between the two time series. An early study showed that all nodes in equatorial regions have large degree (i.e., the number of edges that a node has) regardless of the longitude, whereas only a small fraction of nodes in the mid-latitude regions had large degrees \cite{Tsonis2004PhysicaA}. Climate networks have been further used for understanding mechanisms of climate dynamics and predicting extreme events. For example, early warning signals were constructed from the degree of the nodes and clustering coefficient for climate networks of the Atlantic temperature field~\cite{Vandermheen2013GeophysResLett}. The proposed early warning signals were effective at anticipating the collapse of Atlantic Meridional Overturning Circulation. See section 2.1 of \cite{FanMeng2021PhysRep} for more examples.

\section{Methods for creating networks from correlation matrices\label{sec:math}}

To apply network analysis to correlation matrix data, we need to generate a network from correlation data (usually in the form of a correlation matrix). We call such a network a \emph{correlation network}. Whether correlation network analysis works or is justified depends on this process. Although there are various methods for constructing correlation networks from data, they have pros and cons. Furthermore, there are various unjustified practices around correlation network generation, which may yield serious limitations on the power of correlation network analysis. In this section, we review several major methods.

\subsection{Estimation of covariance and correlation matrices\label{sub:estimation-covariance-matrix}}

How to estimate covariance matrices from observed data when the matrix size is not small is a long-standing question in statistics and surrounding research fields. In particular, the sample covariance matrix, a most natural candidate, is known to be an unreliable estimate of the covariance matrix. See \cite{BickelLevina2008AnnStat-thresholding,Pourahmadi2011StatSci,FanLiaoLiu2016EconometJ,Bun2017PhysRep} for surveys on estimation of covariance matrices. Although the primary focus of this paper is estimation of correlation networks, not covariance or correlation matrices, it is of course important to realize that correlation networks created from unreliably estimated correlation matrices are themselves unreliable. Therefore, we briefly survey a few techniques of covariance and correlation matrix estimation in this section, including providing the notations and preliminaries used in the remainder of this paper. 
This exposition is important in particular because correlation network analysis in non-statistical research fields such as network science and also various applications often ignores statistical perspectives examined in the previous studies.

We denote by $(x_{i\ell})$ an $N\times L$ data matrix, where $N$ is the number of nodes to observe the signal from, and $L$ is the number of samples, which is typically the length of the time series or the number of participants in an experiment or questionnaire. The sample covariance matrix, $C^{\text{sam}} = (C^{\text{sam}}_{ij})$, is given by
\begin{equation}
C^{\text{sam}}_{ij} = \frac{1}{L-1} \sum_{\ell=1}^L (x_{i\ell} - \overline{x}_i) (x_{j\ell} - \overline{x}_j),
\label{eq:sample-covariance-matrix}
\end{equation}
where
\begin{equation}
\overline{x}_i = \sum_{\ell=1}^L \frac{x_{i\ell}}{L}
\end{equation}
is the sample mean of the signal from the $i$th node.\footnote{Note that the $L-1$ in the denominator of Eq.~\eqref{eq:sample-covariance-matrix} is necessary to obtain an unbiased estimator.} Because Eq.~\eqref{eq:sample-covariance-matrix} is a sum of $L$ outer products,
the rank of $C^{\text{sam}}$ is at most $L$. One can understand this fact more easily by rewriting Eq.~\eqref{eq:sample-covariance-matrix} as \begin{equation}
C^{\text{sam}} = \frac{1}{L-1} \sum_{\ell=1}^L \bm{\tilde{x}}_{\ell} \bm{\tilde{x}}_{\ell}^{\top},
\end{equation}
where $\bm{\tilde{x}}_{\ell} =
(x_{1\ell} - \overline{x}_1, \ldots, x_{N\ell} - \overline{x}_N)^{\top}$ 
Therefore, $C^{\text{sam}}$ is singular if $L<N$, while the converse does not hold true in general.

Although covariance matrices are mathematically convenient, they are not normalized. In particular, if we multiply the data from the $i$th node by $c$ ($>0$), then $C_{ij}^{\text{sam}}$ $(= C_{ji}^{\text{sam}})$, where $j\neq i$, changes by a factor of $c$, and $C_{ii}^{\text{sam}}$ changes by a factor of $c^2$, whereas all the other entries of $C^{\text{sam}}$ remain the same. In practice, the data from different nodes may have different baseline fluctuation levels. For example, the price of the $i$th stock may fluctuate much more than that of the $j$th stock because the former has a larger average or the industry to which the $i$th company belongs may be subject to higher temporal variability. The correlation matrix, denoted by $\rho$, normalizes the covariance matrix such that $\rho$ is not subject to the effect of different overall amounts of fluctuations across different nodes. The sample Pearson correlation matrix, denoted by $\rho^{\text{sam}} = (\rho^{\text{sam}}_{ij})$, is defined by
\begin{equation}
\rho^{\text{sam}}_{ij} = 
\frac{\sum_{\ell=1}^L (x_{i\ell} - \overline{x}_i) (x_{j\ell} - \overline{x}_j)}
{\sqrt{\sum_{\ell=1}^L (x_{i\ell} - \overline{x}_i)^2} \sqrt{\sum_{\ell=1}^L (x_{j\ell} - \overline{x}_j)^2}}
=
\frac{C^{\text{sam}}_{ij}} {\sqrt{C^{\text{sam}}_{ii} C^{\text{sam}}_{jj}}}.
\label{eq:rho}
\end{equation}
Note that $\rho^{\text{sam}}_{ii} = 1$ $\forall i \in \{1, \ldots, N\}$. Also note that every sample correlation matrix is a sample covariance matrix of some data but not vice versa.
A correlation matrix is characterized by positive semidefiniteness, symmetry, range of the entries only being $[-1, 1]$, and the diagonal being equal to $1$ \cite{Holmes1991SiamJMatAnalAppl}. The set of full-rank correlation matrices for a fixed $N$ is called the elliptope, which has its own geometric structure \cite{Tropp2018DiscreteComputGeom,Thanwerdas2021Lncs}.
For standardized samples $y_{i\ell} = (x_{i\ell} - \overline{x}_i)/\sqrt{C^{\text{sam}}_{ii}}$, the Euclidean distance between vectors
$(y_{i1}, \ldots, y_{iL})$ and $(y_{j1}, \ldots, y_{jL})$ is given by
\begin{equation}
d_{ij}^2 = \sum_{\ell=1}^L \left( y_{i\ell} - y_{j\ell} \right)^2 = 2 - 2 \rho_{ij}^{\text{sam}}.
\label{eq:def-d_{ij}}
\end{equation}
Therefore, given the sample correlation matrix, $d_{ij} = \sqrt{2(1-\rho_{ij}^{\text{sam}})}$ defines a Euclidean distance \cite{Mantegna1999EurPhysJB,Mantegna1999book}.

In the following, we refer to covariance matrices in some cases and correlation matrices in others, often interchangeably. This is because the correlation matrix, which is a normalized quantity, should be used in most data analyses, while the covariance matrix allows better mathematical analysis in most cases. This convention is not problematic because, if we standardize the given data first and then calculate the covariance matrix for the standardized data ($y_{i\ell}$), then the obtained covariance matrix is also a correlation matrix. Therefore, the mathematical results for covariance matrices also hold true for correlation matrices as long as we feed the pre-standardized data to the analysis pipeline.

With the above consideration in mind, we now stress that it is important to distinguish the \emph{sample} covariance matrix, which is calculated from empirical data, from the theoretical or `true' (also called \emph{population}) covariance matrix. One may use the true covariance matrix to model the observed data mathematically in terms of a random process described by a (stationary) probability distribution. Let $X_i$ denote a random variable for $i \in 1,\ldots,N$. The true covariance matrix $C=(C_{ij})$ is given by
\begin{equation}
C_{ij} = E[(X_i - \mu_i)(X_j - \mu_j)],
\label{eq:def-covariance-matrix}
\end{equation}
where $E$ represents the expectation, and $\mu_i = E[X_i]$ is the ensemble mean of $X_i$. Equation~\eqref{eq:def-covariance-matrix} implies that a covariance matrix is a symmetric matrix. It is also a positive semidefinite matrix. Conversely, a symmetric positive semidefinite matrix $C$ is always a covariance matrix for the following reason. Any positive semidefinite matrix $M$ allows a positive semidefinite matrix, denoted by $M^{\frac{1}{2}}$, as square root. We set 
\begin{equation}
\begin{pmatrix}
X_1\\ \vdots \\ X_N
\end{pmatrix} = C^{\frac{1}{2}} \begin{pmatrix}
U_1\\ \vdots \\ U_N
\end{pmatrix},
\end{equation}
where $U_1$, $\ldots$, $U_N$ are independent random variables, with each having mean $0$ and variance $1$.
Then, it is straightforward to see that
$(X_1, \ldots, X_N)^{\top}$ has mean $0$ and covariance matrix $C$.

Having distinguished population and sample covariance matrices, we now look for a relationship between the two. If we regard $X_i$ as a random variable, and the observed data as a possible realization of that variable, then we must also regard the sample covariance matrix $C^{\text{sam}}$ as a random variable, and the empirical sample correlation matrix as a single realization of that variable. Then, the relevant question is the characterization of the probability distribution governing the sample covariance matrix, given the true covariance matrix $C$ which acts a set of fixed parameters (to be estimated from the data) for the distribution. Clearly, the number $L$ of observations, regarded as the number of independent draws for each random variable $X_i$, is another parameter (not to be estimated). For any finite $L$, the sample covariance matrix obeys the so-called Wishart distribution with $L$ degrees of freedom, denoted by $W_N(L, C)$, under the assumption that the $L$ samples are i.i.d.\,and obey the multivariate normal distribution whose covariance matrix is $C$~\cite{Whittaker1990book,Lauritzen1996book,Anderson2003book,Bun2017PhysRep}. We obtain $E[C^{\text{sam}}] = C$. In other words, the sample covariance matrix is an unbiased estimator of the true covariance matrix, called the Pearson estimator in statistics. The variance of $C^{\text{sam}}_{ij}$ is equal to
$(C_{ij}^2 + C_{ii} C_{jj})/L$. In fact, $C^{\text{sam}}$ is a problematic 
substitute of $C$, and the use of $C^{\text{sam}}$ in applications in place of $C$ tends to fail; see \cite{Bun2017PhysRep} for an example in portfolio optimization. An intuitive reason why $C^{\text{sam}}$ is problematic is that, if $L$ is not much larger than $N$, which is often the case in practice, one would need to estimate many parameters from a relatively few observations. Specifically, the covariance and correlation matrices have $N(N+1)/2$ and $N(N-1)/2$ unknowns to infer, respectively, whereas there are $L$ samples of vector $(x_{1\ell}, \ldots, x_{N\ell})$ available \cite{Dempster1972Biometrics}. If $N/L$ is not vanishingly small (called the large dimension limit or the Kolmogorov regime \cite{Bun2017PhysRep}), then the estimation would fail.
As an extreme example, if $L < N$, then $C^{\text{sam}}$ is singular, but the true $C$ may be nonsingular.
Even if $L\ge N$, matrix $C^{\text{sam}}$ may be ill-conditioned if $L$ is not sufficiently greater than $N$, whereas $C$ may be well-conditioned.

Therefore, covariance selection to reduce the number of parameters to be estimated is a recommended practice when $L$ is not large relative to $N$ \cite{Dempster1972Biometrics}. One also says that we impose some structure on the estimator of the covariance matrix, with the mere use of the sample covariance matrix as an estimate of the true covariance matrix corresponding to no assumed structure.

A major method of covariance selection is to impose sparsity on the covariance matrix or the so-called precision matrix (also called the concentration matrix), which is the inverse of the covariance matrix (with entries representing un-normalized partial correlation coefficients; see section~\ref{sub:partial-corr}). Note that a sparse precision matrix does not imply that the corresponding covariance matrix (i.e., its inverse) is sparse. Graphical lasso (see section~\ref{sub:graphical-lasso}) is a popular method to estimate a sparse precision matrix. Another major method to estimate a sparse correlation matrix is to threshold on the value of the correlation to discard node pairs with correlation values close to $0$ (see section~\ref{sub:dichotomizing}). Another common method of covariance selection, apart from estimating a sparse covariance matrix, is covariance shrinkage (see \cite{Ledoit2022JFinEconomet} for a review). With covariance shrinkage, the estimated covariance matrix is a linear weighted sum of the sample covariance matrix, $C^{\text{sam}}$, and a much simpler matrix, called the shrinkage target, such as the identity matrix \cite{Ledoit2003JEmpirFin,Ledoit2004JMultivarAnal,Chen2010IeeeTransSignalProc} or the so-called single-index model (which is a one-factor model in factor analysis terminology and is an approximation of $C^{\text{sam}}$ by a rank-one matrix plus residuals) \cite{Ledoit2003JEmpirFin}. Note that the shrinkage target is a biased estimator of $C$. These and other covariance selection methods balance between the estimation biases and variances.

An advanced estimation method for the entire correlation matrix, based on RMT, is the so-called optimal rotationally invariant estimator (RIE)~\cite{Bun2017PhysRep,ibanez2023noise}. Roughly speaking, the RIE is the closest (in some spectral sense) matrix, among all those having the same eigenvectors as the sample correlation matrix, to the `true’ correlation matrix. It uses a certain self-averaging property to infer the spectrum of the true matrix from that of the sample matrix, and then to compute the spectral distance from the true matrix to be minimized~\cite{Bun2017PhysRep}. A specific cross-validated version of the RIE has been recently shown to outperform several other estimators~\cite{ibanez2023noise}. Since the RIE requires some notion of RMT to be properly defined, we discuss it in section~\ref{sec:RMT}.

\subsection{Dichotomization\label{sub:dichotomizing}}

In this and the following subsections, we present several methods to generate undirected networks from correlation matrices.

\begin{figure}
\begin{center}
\includegraphics[width=16cm]{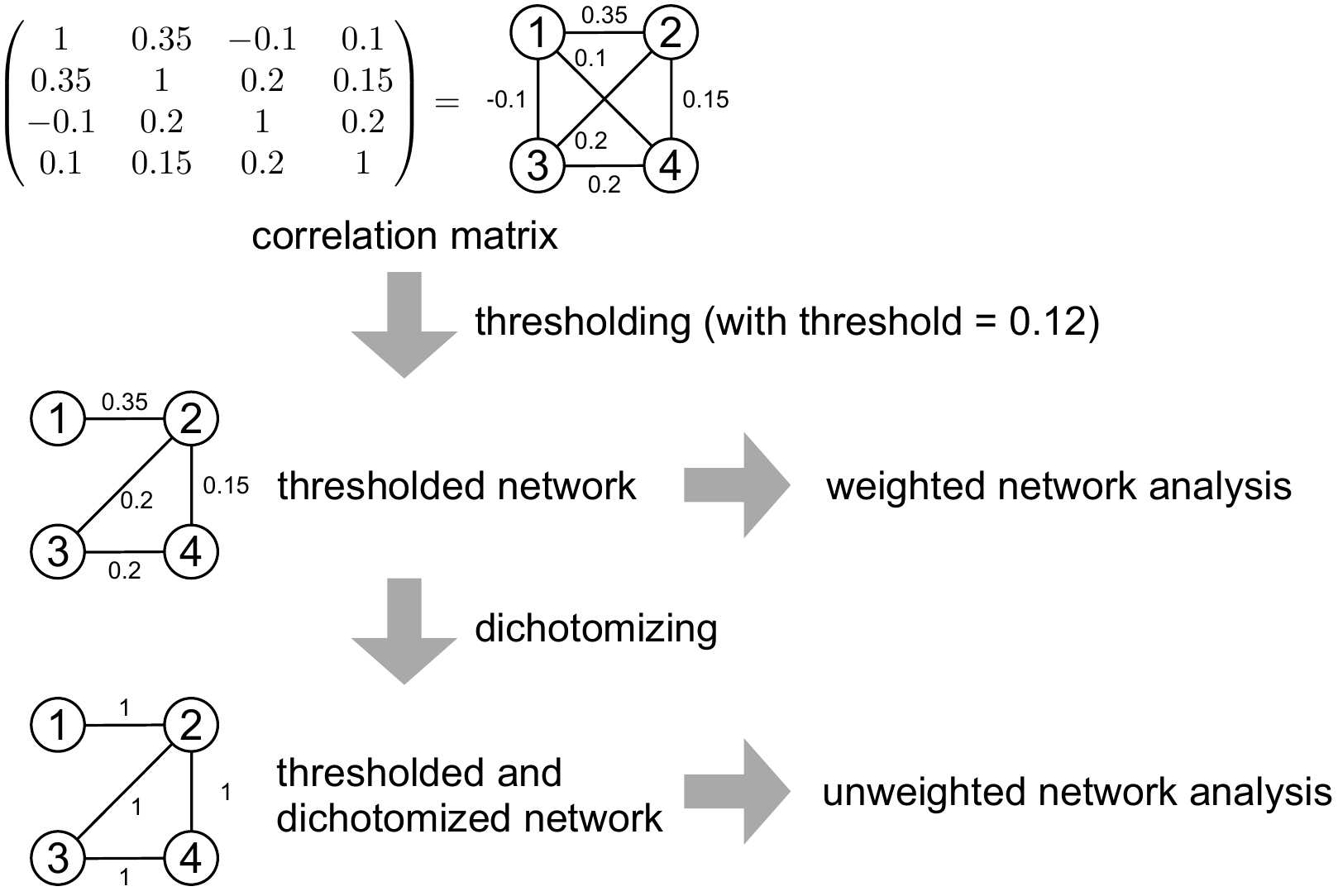}
\caption{Thresholding a correlation matrix. We set the threshold at $\theta = 0.12$. If we only threshold the correlation matrix, we obtain a weighted network. If we further dichotomize the thresholded matrix, we obtain an unweighted network. A different $\theta$ yields a different network in general.}
%
%
\label{fig:thresholding}
\end{center}
\end{figure}

A simple method to generate a network from the given correlation matrix is thresholding, which in its simplest form entails setting a threshold $\theta$, and placing an unweighted edge $(i, j)$ if and only if the Pearson correlation $\rho_{ij} \ge \theta$; otherwise, we do not include an edge $(i, j)$. It is also often the case that one thresholds on $\left| \rho_{ij} \right|$. There are mainly two choices after the thresholding. First, we may discard the weight of the surviving edges to force it to $1$, creating an unweighted network. Second, we may keep the weight of the surviving edge to create a weighted network.
See Fig.~\ref{fig:thresholding} for these two cases. The literature often use the term thresholding in one of the two meanings without clarification. In the remainder of this paper, we call the first case \emph{dichotomizing} (which can be also called binarizing), which is, precisely speaking, a shorthand for ``thresholding followed by dichotomizing''. We discuss dichotomized networks in this section and threshold networks without dichotomization (yielding weighted networks) in sections~\ref{sub:weighted} and \ref{sub:graphical-lasso}.

Dichotomizing has commonly been used across research areas. However, researchers have repeatedly pointed out that dichotomizing is not recommended for multiple reasons.

First, no consensus exists regarding the method for choosing the threshold value \cite{Rubinov2011Neuroimage, LeeKang2012IeeeTransMedImaging, Devicofallani2014PhilTransRSocB,Garrison2015Neuroimage,Devicofallani2017PlosComputBiol,ChungLee2019NetwNeurosci} despite that results of correlation network analysis are often sensitive to the threshold value \cite{Cole2010Neuroimage, LeeKang2012IeeeTransMedImaging, Scheinost2012Neuroimage,Garrison2015Neuroimage,Devicofallani2017PlosComputBiol, ChungLee2019NetwNeurosci, Peel2022NatComm}. In a related vein, if a single threshold value is applied to the correlation matrix obtained from different participants in an experiment, which is typical in neuroimaging data analysis and referred to as an absolute threshold \cite{Garrison2015Neuroimage,Vandenheuvel2017Neuroimage}, the edge density can vary greatly across participants. Since edge density is heavily correlated with many network measures, this can be seen as introducing a confound into subsequent analyses and casts doubt on consequent conclusions, e.g., that sick participants tend to have less 
small-world brain networks than healthy controls. (In this example, a network with a large edge density would in general yield a small average path length and large clustering coefficient, leading to the small-world property, so that density differences alone could have driven the observed effect.) An alternative method for setting the threshold is the so-called proportional thresholding, with which one keeps a fixed fraction of the strongest (i.e., most correlated) edges to create a network, separately for each participant \cite{Garrison2015Neuroimage,Vandenheuvel2017Neuroimage}; also see  \cite{Achard2007PlosComputBiol,Vandenheuvel2008Neuroimage,WangWang2009HumanBrainMapping} for early studies. In this manner, the thresholded networks for different participants have the same density of edges. However, while the proportional thresholding may sound reasonable, it has its own problems \cite{Vandenheuvel2017Neuroimage}. First, because different participants have different magnitudes of overall correlation coefficient values, the proportional threshold implies that one includes relatively weakly correlated node pairs as edges for participants with an overall low correlation coefficients. This procedure increases the probability of including relatively spurious node pairs, which can be regarded as type I errors (i.e., false positives), increasing noise in the resulting network. (Also see \cite{Vanwijk2010PlosOne,Alexanderbloch2010FrontSystNeurosci} for discussion on this matter.) Second, the overall correlation strength is often predictive of, for example, a disease in question. The proportional threshold enforces the same edge density for the different participants' networks. Therefore, it gives up the possibility of using the edge density, which is a simplest network index, to account for the group difference. If one uses the absolute threshold, the edge density is different among participants, and one can use it to characterize participants. The edge density in the proportional thresholding is also an arbitrary parameter.

Second, apart from false positives due to keeping small-correlation node pairs as edges, correlation networks at least in its original form suffer from false positives because pairwise correlation does not differentiate between direct effects (i.e., nodes $i$ and $j$ are correlated because they directly interact) and indirect effects (i.e., nodes $i$ and $j$ are correlated because $i$ and $k$ interact and $j$ and $k$ interact). In other words, correlations are transitive. The correlation coefficient is lower-bounded by \cite{Langford2001AmStat}
\begin{equation}
\rho_{ij} \ge \rho_{ik} \rho_{jk} - \sqrt{(1-\rho_{ik})^2(1-\rho_{jk})^2}.
\label{eq:Langford-bound}
\end{equation}
Equation~\eqref{eq:Langford-bound} implies that if $\rho_{ik}$ and $\rho_{jk}$ are large, i.e., sufficiently close to $1$, then $\rho_{ij}$ is positive.
Furthermore, this lower bound of $\rho_{ij}$ is usually not tight, suggesting that $\rho_{ij}$ tends to be more positive than what Eq.~\eqref{eq:Langford-bound} suggests when $\rho_{ik}, \rho_{jk} > 0$
\cite{Gillis2011Bioinfo,Zalesky2012Neuroimage}. 
This false positive problem is the main motivation behind the definition of the partial correlation networks and related methods with which to remove such a third-party effect, i.e., influence of node $k$ in  Eq.~\eqref{eq:Langford-bound}. (See section~\ref{sub:partial-corr}.) Instead, one may want to suppress false positives by carefully choosing a threshold value. Let us consider the absolute thresholding. For example, if $i$ and $j$ do not directly interact, $i$ and $k$ do, $j$ and $k$ also do, yielding $\rho_{ij} = 0.4$, $\rho_{ik} = 0.7$ and $\rho_{jk} = 0.6$, then setting $\theta=0.5$ enables us to remove the indirect effect by $k$. However, it may be the case that $i'$ and $j'$ do not directly interact, $i'$ and $k'$ do, $j'$ and $k'$ also do, yielding $\rho_{i' j'} = 0.2$, $\rho_{i' k'} = 0.3$, and $\rho_{j' k'} = 0.4$. Then, thresholding with $\theta=0.5$ dismisses direct as well as indirect interactions (that is, it introduces false negatives). A related artifact introduced by the combination of thresholding and indirect effects is that thresholding tends to inflate the abundance of triangles, as measured by the clustering coefficient for dichotomized (and therefore unweighted) networks, and other short cycles~\cite{Steuer2006BriefBioinfo,Zalesky2012Neuroimage}; even correlation networks generated by dichotomizing randomly and independently generated data $\{x_{i\ell} \}$ have high clustering coefficients \cite{Zalesky2012Neuroimage}. This phenomenon resembles the fact that spatial networks tend to have high clustering just because the network is spatially embedded~\cite{Barthelemy2011PhysRep,Barthelemy2018book}.

Third, whereas thresholding has been suggested to be able to mitigate uncertainty on weak links (including the case of the proportional thresholding to some extent) and enhance interpretability of the graph-theoretical results (e.g., \cite{Devicofallani2014PhilTransRSocB}),
%
%
thresholding in fact discards the information contained in the values of the correlation coefficient. For example, in Fig.~\ref{fig:thresholding}, thresholding turns a correlation of $-0.1$ and $0.1$ into the absence of an edge. Furthermore, if we dichotomize the edges that have survived thresholding, a correlation of $0.2$ and $0.35$ are both turned into the presence of an edge.

There are various methods to try to mitigate some of these problems. In the remainder of this section, we cover methods related to dichotomizing.

One family of solutions is to integrate network analysis results obtained with different threshold values \cite{Devicofallani2014PhilTransRSocB} (but see \cite{Langer2013PlosOne} for a critical discussion).
For example, one can calculate a network index, such as the node's degree, denoted by $\alpha$, as a function of the threshold value, $\theta$, and fit a functional form to the obtained function $\alpha(\theta)$ to characterize the node \cite{Scheinost2012Neuroimage}.
Similarly, one can calculate $\alpha$ for a range of $\theta$ values and take an average of $\alpha$ \cite{Lynall2010JNeurosci,Ginestet2011PlosOne}.
In the case of group-to-group comparison, an option along this line of idea is the functional data analysis (FDA), with which one looks at $\alpha$ as a function of $\theta$ across a range of $\theta$ values and statistically test the difference between the obtained function for different groups by a nonparametric permutation test \cite{Bassett2012Neuroimage,Hosseini2012PlosOne}.
In these methods, how to choose the range of $\theta$ is a nontrivial question.

A different strategy is to determine the threshold value according to an optimization criterion. For example, a method was proposed
\cite{Devicofallani2017PlosComputBiol} for determining the threshold value as a solution of the optimization of the trade-off between the efficiency of the network \cite{Latora2001PhysRevLett} and the density of edges.
Another method to set $\theta$ is to use the highest possible threshold that guarantees all or most (e.g., 99\%) of nodes are connected \cite{Bassett2006PNAS}.

%

The so-called maximal spanning tree is an easy and classical method to automatically set the threshold by guaranteeing that the network is connected \cite{Mantegna1999EurPhysJB,Mantegna1999book}, while at the same time avoiding the creation of edges that would form loops (and are therefore unnecessary for connectedness). One adds the largest correlation node pairs as edges one by one under the condition that the generated network is the tree. In the end, the maximal spanning tree contains all the $N$ nodes, and the number of edges is $N-1$. Thanks to the mapping from (large) sample correlation coefficients to (small) Euclidean distances established by Eq.~\eqref{eq:def-d_{ij}}, the maximal (in the sense of correlation) spanning tree is sometimes called the \emph{minimum} (in the sense of distance) spanning tree~\cite{Mantegna1999book,Mantegna1999EurPhysJB} (MST). The MST can be viewed as the graph achieving the overall minimum length  among all graphs that make the $N$ nodes reachable from one another. Here, the minimum length is defined as the sum of the lengths of its realized edges, and the length of an edge is the metric distance between its endpoints. The maximal spanning tree also allows a hierarchical tree representation, which facilitates interpretation \cite{Mantegna1999EurPhysJB,Vandewalle2001QuantFinance,Onnela2002EurPhysJB}. However, the generated network is extreme in the sense that it is a most sparse network among all the connected networks on $N$ nodes, without any triangles. A variant of the maximum spanning tree is to sequentially add edges with the largest correlation value under the constraint that the generated network can be embedded on a surface of a prescribed genus value (roughly speaking, the given number of holes) without edge intersection \cite{Tumminello2005PNAS}. If the genus is constrained to be zero, the resulting network is a planar graph, called the planar maximally filtered graph (PMFG). The PMFG contains $3N-6$ edges. The PMFG contains more information than the maximum spanning tree (which is in any case contained in the PMFG), such as some cycles and their statistics. Note that these methods effectively produce a threshold for the correlation value to be retained, but on the other hand preserve only some of the edges that exceed the threshold. Indeed, they are (designed to be) irreducible to the mere identification of an overall threshold value, with their merit residing in the introduction of higher-order geometric constraints guiding the dichotomization procedure.

The maximal spanning forest (MSF) is similar to the maximal spanning tree but extracts different information from correlation matrix data \cite{Mastrandrea2017SciRep}. In the MSF, each node has out-degree 1 with the out-going edge from node $v$ pointing to the node that is the most strongly correlated with $v$. A node may then have no in-coming edge (i.e., in-degree equal to $0$) or in-degree larger than $1$. The MSF is a directed network with $N$ (directed) edges and without cycles of length greater than two (reciprocated edges may exist where two nodes are most strongly correlated with each other). The MSF is in general not (even weakly) connected, and the different ``chain-like'' components may be of practical interest \cite{Mastrandrea2017SciRep}. In contrast, the MST is composed of a single connected component. Indeed, if one adds edges to connect the components of the MSF into a single component by considering all node pairs between different components and sequentially adding the highest-correlation edges only when the added edge connects two components into one, the undirected version of this single connected component is an MST~\cite{Mastrandrea2017SciRep}.

Another method related to the maximum spanning tree is to use the $k$ nearest neighbor graph of the correlation matrix, in which each $i$th node is connected at least to the $k$ nodes with the highest correlation with the $i$th node \cite{Alexanderbloch2010FrontSystNeurosci}. Yet another choice, which is designed for general weighted networks, is the disparity filter, with which one preserves only statistically significant edges to generate the \emph{network backbone} \cite{Serrano2009PNAS,gemmetto2017irreducible}. Note that, with these methods as well, some lower-correlation node pairs are retained as edges and some higher-correlation edges are discarded.

\textit{Application example}: Wang et al.\ compared functional networks of the brain between children with attention-deficit hyperactivity disorder (ADHD) and healthy controls using fMRI data \cite{WangZhuHe2009HumanBrainMapping}. The brain scan lasted for eight minutes in the resting state, and the brain image was acquired every two seconds. 
The authors used a previously published popular brain atlas defined by three-dimensional coordinates of 90 anatomical regions of interest (45 per hemisphere), each of which defines a node of the network. After various steps of preprocessing the original data, they computed the Pearson correlation coefficient between the time series at each pair of nodes, separately for each child. Aware of the problem inherent in choosing a threshold value, which we discussed earlier in this section, the authors examined dichotomized networks for a range of threshold values $\theta \in [0.05, 0.5]$. As downstream analysis, they measured the small-world-ness of networks in terms of the so-called global efficiency and local efficiency \cite{Latora2001PhysRevLett}. A large global efficiency value and a small local efficiency value suggest that the network has a small-world property. They found that, while the networks of both ADHD and healthy control children were small-world, those of children with ADHD were somewhat less small-world than those of the controls across a wide range of $\theta$; the difference was statistically significant for the local efficiency in some range of $\theta$.

\subsection{Persistent homology\label{sub:persistent-homology}}

In the previous section, we discussed the idea of integrating analysis of dichotomized networks over different threshold values to mitigate the effect of selecting a single threshold value.
Topological data analysis, or more specifically, persistent homology, provides a systematic and mathematically founded suite of methods to do such analyses. In topological analysis in general, one focuses on properties of shapes that do not change under continuous deformations of them, such as stretching and bending. Such topologically invariant properties include the numbers of connected components and of holes, which can be calculated through the so-called homology group.
Persistent homology captures the changes 
in the network structure over multiple scales, or over a range of threshold values, clarifying topological features that are robust with respect to the threshold choice. In this broader perspective, networks are only a particular instance of the type of topological object under major consideration in topological data analysis, called simplicial complexes, because networks only consider pairwise interactions. One may want to consider the clique complex, in which each $k$-clique (i.e., complete subgraph with $k$ nodes) in the network is defined to be a higher-order object called a $(k-1)$-simplex and belongs to the simplicial complex of the given data. Note that the clique complex contains all the edges of the original network as well because an edge is a 2-clique by definition.
For reviews of topological data analysis including persistent homology, see \cite{Ghrist2008BullAmerMathSoc, Horak2009JStatMech, Otter2017EpjDataSci, Aktas2019ApplNetwSci, Sizemore2019NetwNeurosci}.

To analyze correlation matrix data using persistent homology, we start with a point cloud, with each point corresponding to a node in a correlation network. Then, we introduce a distance between each pair of nodes, $d_{ij}$, where $1\le i, j\le N$. By setting a threshold value $\theta'$, we obtain a dichotomized network, or, e.g., a clique complex, depending on the choice, denoted by $G_{\theta'}$. The simplicial complexes with varying $\theta'$ values forms a filtration, i.e., with the nestedness property
\begin{equation}
G_{\theta'_1} \subseteq G_{\theta'_2} \subseteq G_{\theta'_3}\subseteq \cdots,
\mbox{ for } \theta'_1 \le \theta'_2 \le \theta'_3 \le \cdots, 
\label{eq:filtration}
\end{equation}
with the inclusion relationship in Eq.~\eqref{eq:filtration} referring to that of the edge set and that of any higher-order simplex object. The collection of clique complexes in Eq.~\eqref{eq:filtration} is called the Vietoris-Rips filtration. Simply put, with a larger threshold on the distance between nodes, the generated simplicial complex has more edges (and higher-order simplexes in the case of the clique complex). In practice, it suffices to consider the sequence of threshold values $\{ \theta'_1, \theta'_2, \ldots \}$, such that $G_{\theta'_{m+1}}$, contains just one more edge (and some higher-order simplexes containing that edge) than $G_{\theta'_m}$; if there are multiple node pairs with exactly the same distance value, the corresponding multiple edges not in $G_{\theta'_{m}}$ simultaneously appear in $G_{\theta'_{m+1}}$. If $\theta'$ is taken to vary over all possible topologies, the simplicial complex $G_{\theta'_1}$ is composed of $N$ isolated nodes, while the last one in the nested sequence is the complete graph (that is, precisely, the corresponding clique complex) of $N$ nodes.

A large correlation should correspond to a small $d_{ij}$. One can realize this by setting $d_{ij} = f(\rho_{ij}^{\text{sam}})$, where $f$ is
a monotonically decreasing function. Then, the network interpretation of Eq.~\eqref{eq:filtration} simply states the nested relationship in which the edges existing in a dichotomized network are included in dichotomized networks with smaller thresholds. However, while this practice is common, $d_{ij}$ is not mathematically guaranteed to be a distance metric, typically violating the triangle inequality if we use an arbitrary monotonically decreasing function $f$. Therefore, one often uses $f$ that makes $d_{ij}$ a distance metric, such as
Eq.~\eqref{eq:def-d_{ij}} or variants thereof. Then, the ensuing topological data analysis of correlation networks is underpinned by stronger mathematical foundations.

The next step is to calculate for each $G_{\theta'}$ the homology groups or associated quantities, such as the $k$th Betti number.
The zeroth and first Betti numbers are the numbers of connected components and essentially different cycles, respectively. These and other topological features of $G_{\theta'}$ depend on $\theta'$. For example, each connected component and cycle may appear (through closing loops) and disappear (through mergers with others) as one gradually increases the distance threshold $\theta'$ from $\theta' = 0$, at which all the nodes are isolated. One can precisely visualize the occurrence of the birth and death events of each component by the persistence barcodes or persistence diagrams. For example, the persistence diagram represents each connected component, cycles, two-dimensional voids, etc.\ as a point $(x, y)$ in the two-dimensional space in which $x$ represents the $\theta'$ value at which, e.g., a cycle appears and $y$ ($\ge x$) represents the $\theta'$ value at which the same cycle disappears. If $y-x$ is large, that particular feature of the data is robust over changing scales because it is independent of the specific value of $\theta'$ within a relatively large range.

We are often motivated to compare different correlation matrices or networks, such as in the comparison of functional brain networks between the patient group and control group. Quantitative comparison between two persistent diagrams provides a threshold-free method to effectively compare dichotomized networks. A persistence diagram consists of a set of two-dimensional points $(x, y)$. The bottleneck and Wasserstein distance, between persistence diagrams $P_1$ and $P_2$, which are commonly used, first consider the best matching between $(x, y)$ in $P_1$ and that in $P_2$. For the obtained best matching, both distance metrics measure the distance between $(x, y)$ in $P_1$ and that in $P_2$ in the Euclidean space and tally up the distance over all the points in different manners. For instance, the bottleneck distance is given by
$d(P_1, P_2) = \inf_{\gamma} \sup_{(x, y) \in P_1} \left\| (x, y) - \gamma((x, y))\right\|_{\infty}$, where $\gamma$ is a matching between $P_1$ and $P_2$.

Persistent homology has been applied to correlation networks in neuronal activity data \cite{LeeKang2012IeeeTransMedImaging, Petri2014JRSocInterface, Giusti2015Pnas, ChungLee2019NetwNeurosci}, gene co-expression data \cite{Otter2017EpjDataSci, Duman2018IntJGenomics, Shnier2019JRSocInterface}, financial data \cite{Leibon2008Pnas, Gidea2017Netscix}, and to co-occurrence networks of characters in literature work and in academic collaboration \cite{Rieck2018IeeeTransVisComputGraphics}. Scalability of persistent homology algorithms remains a concern. However, it may be of less concern for correlation network analysis because the number of nodes allowed for correlation networks is typically limited by the length of the data, not by the speed of algorithms (see section~\ref{sub:estimation-covariance-matrix}).

\subsection{Weighted networks\label{sub:weighted}}

A strategy for avoiding the arbitrariness in the choice of the threshold value and loss of information in dichotomizing is to use weighted networks, retaining the pairwise correlation value as the edge weight \cite{Langfelder2008Bioinfo,Rubinov2011Neuroimage}. Although there are numerous settings in network science where negative edge weights are considered, they are generally more difficult to treat. (See section~\ref{sub:negative}.) As such, two common methods to create positively weighted networks are (1) using the absolute value of the correlation coefficient as the edge weight and (2) ignoring negatively weighted edges and only using the positively weighted edges. Both methods dismiss some information contained in the original correlation matrix, i.e., the sign of the correlation or the magnitude of the negative pairwise correlation. Nonetheless, these transformations are widely used because many methods are available for analyzing general positively weighted networks, many of which are extensions of the corresponding methods for unweighted networks. One can also use methods that are specifically designed for weighted networks
\cite{Horvath2011book}. 

It should be noted that weighted networks share the problem of false positives due to indirect interaction between nodes with the unweighted networks created by dichotomization. We also note that, in contrast to thresholding (which may be followed by dichotomization), node pairs with any small correlation (i.e., correlation coefficient close to $0$) are kept as edges in the case of the weighted network. This may increase the uncertainty of the generated network and hence of the subsequent network analysis results.

Thresholding operations in statistics literature to increase the sparsity of the estimated covariance matrix often produce weighted networks. This is in contrast to the dichotomization, which produces unweighted networks. Hard thresholding in statistics literature refers to coercing $C^{\text{sam}}_{ij}$, with $i\neq j$, to $0$ if $\left| C^{\text{sam}}_{ij} \right| < \theta$ and keep the original
$C^{\text{sam}}_{ij}$ if $\left| C^{\text{sam}}_{ij} \right| \ge \theta$
\cite{Donoho1994Biometrika,BickelLevina2008AnnStat-thresholding,Elkaroui2008AnnStat-thresholding,Rothman2009JAmStatAssoc}. Soft thresholding \cite{Donoho1994Biometrika,Tibshirani1996JRStatistSocB,Rothman2009JAmStatAssoc} transforms $C^{\text{sam}}_{ij}$ by a continuous non-decreasing function of $C^{\text{sam}}_{ij}$, denoted by $f(C^{\text{sam}}_{ij})$, such that 
\begin{equation}
f(x) =
\begin{cases}
x - \theta & \text{if } x \ge \theta,\\
0 & \text{if } -\theta < x < \theta,\\
x + \theta & \text{if } x \le - \theta.
\end{cases}
\label{eq:soft-thresholding}
\end{equation}
This assumption implies that, in contrast to hard thresholding, there is no discontinuous jump in the transformed edge weight at $C^{\text{sam}}_{ij} = \pm \theta$.
Both hard and soft thresholding, as well as a more generalized class of thresholding function $f(x)$ \cite{Rothman2009JAmStatAssoc}, do not imply dichotomization and therefore generate weighted networks. In numerical simulations, all these thresholding methods to generate weighted networks outperformed the sample covariance matrix in estimating true sparse covariance matrices~\cite{Rothman2009JAmStatAssoc}. The same study also found that there was no clear winner between hard or soft thresholding, while combination of them tended to perform somewhat better than other types of thresholding. 

Adaptive thresholding refers to using threshold values that depend on $(i, j)$. An example is to use Eq.~\eqref{eq:soft-thresholding}, but with an $(i, j)$-dependent threshold value, denoted by $\theta_{ij}$, in place of $\theta$, with $\theta_{ij} = \overline{c} \sqrt{\text{Var}_{ij} \cdot \log N / L}$, where $\overline{c}$ is a constant, and $\text{Var}_{ij}$ is an estimate of the variance of $(X_i - \mu_i)(X_j - \mu_j)$
~\cite{CaiLiu2011JAmStatAssoc}.
This adaptive thresholding theoretically converges faster to the real covariance matrix and performs better numerically than universal thresholding schemes (i.e., using a threshold value independent of $(i, j)$).  

Tapering estimators also threshold and suppress $C^{\text{sam}}_{ij}$ in an $(i, j)$-dependent manner. A tapering estimator for the $(i, j)$ entry of the covariance matrix is given by $\overline{f}(i, j) \cdot \frac{L-1}{L} C^{\text{sam}}_{ij}$, where
$\overline{f}(i, j) = 1$ if $|i-j| \le k/2$, $\overline{f}(i, j) = 2 - 2|i-j|/k$ if $k/2 < | i - j | < k$, and $\overline{f}(i, j) = 0$ if $| i - j | \ge k$. Here, tapering parameter $k$ is an even integer, and $\frac{L-1}{L} C^{\text{sam}}_{ij}$ is the maximum likelihood estimator of $C_{ij}$. The tapering estimator more strongly suppresses the $(i, j)$ entries that are farther from the diagonal respecting the sparsity of the estimated covariance matrix. This estimator is optimal in terms of the rate of convergence to the true covariance matrix, and the value of $k$ realizing the optimal rate differs between the two major matrix norms with which the estimation error is measured \cite{CaiZhangZhou2010AnnStat}.

\textit{Application example}: Chen et al.\ compared gene co-expression networks between people with schizophrenia and non-schizophrenic controls \cite{ChenChengGrennan2013MolPsychiatry}. They used the Weighted Correlation Network Analysis (WGCNA) software (see section~\ref{sec:software}), which is frequently used in gene co-expression network analysis, to construct weighted networks of genes. In short, it uses $\left| \rho_{ij} \right|^{\beta}$ as the weight of edge $(i, j)$, where $\beta$ is a parameter. Then, they compared gene modules, or communities of the network, between the schizophrenia and control groups. The module detection was carried out by an algorithm implemented in WGCNA. There was no significant difference between the schizophrenia and control groups in terms of the module structure (i.e., which genes are in each module). However, the eigengene --- that is, the first principal component of the co-expression matrix within a module --- of each of two modules was significantly associated with schizophrenia compared to the control. The hub genes of each of these two modules, \textit{NOTCH2} and \textit{MT1X}, were the largest contributor to the respective eigengenes. The authors further carried out biological analyses of these two genes to clarify where their expressions are upregulated and the functions in which these genes may be involved. The results were similar for a biopolar disorder data set.

\subsection{Negative weights\label{sub:negative}}

Correlation matrices have negative entries in general. In the case of both unweighted and weighted correlation networks, we often prohibit negative edges either by coercing negative entries of the correlation matrix to zero or by taking the absolute value of the pairwise correlation before transforming the correlation matrix into a network. We prohibit negative edges for two main reasons. First, in some research areas, it is often difficult to interpret negative edges. In the case of multivariate financial time series, a negative edge implies that the price of the two assets tend to move in the opposite manner, which is not difficult to interpret. In contrast, when fMRI time series from two brain regions are negatively correlated, it does not necessarily imply that these regions are connected by inhibitory synapses, and it is not straightforward to interpret negative edges in brain dynamics data~\cite{Vandijk2010JNeurophysiol,Devicofallani2014PhilTransRSocB}.
%
%
Second, compared to weighted networks and directed networks, we do not have many established tools for analyzing networks in which positive and negative edges are mixed, i.e., signed networks. Signed network analysis is still emerging \cite{Tang2016AcmComputSurv} 

In fact, negative edges may provide useful information. For example, they benefit community detection because, while many positive edges should be within a community, negative edges might ideally connect different communities rather than lie within a community. Some community detection algorithms for signed networks exploit this idea~\cite{Tang2016AcmComputSurv,Zhan2017JComparativeNeurol}. Another strategy for analyzing signed network data is to separately analyze the network composed of positive edges and that composed of negative edges and then combine the information obtained from the two analyses. For example, the modularity, an objective function to be maximized for community detection, can be separately defined for the positive network and the negative network originating from a single signed network and then combined to define a composite modularity to be maximized~\cite{Gomez2009PhysRevE,Rubinov2011Neuroimage}. While these methods are designed for general signed networks, they have been applied to brain correlation networks \cite{Rubinov2011Neuroimage,Zhan2017JComparativeNeurol}.

Another type of approach to signed weighted networks is nonparametric weighted stochastic block models~\cite{Aicher2015JCompNetw,Peixoto2018PhysRevE}, which are useful for modeling correlation matrix data. Crucially, this method separately estimates the unweighted network structure and the weight of each edge but in a unified Bayesian framework. By imposing a maximum-entropy principle with a fixed mean and variance on the edge weight, they assumed a normal distribution for the signed edge weight. Because the edge weight in the case of correlation matrices, i.e., the correlation coefficient, is confined between $-1$ and $1$, an ad-hoc transformation to map $(-1, 1)$ to $(-\infty, \infty)$ such as $y = 2 \arctanh x = \ln \frac{1+x}{1-x}$ is applied before fitting the model. One can assess the goodness of such an ad-hoc transformation by a posteriori comparison with different forms of functions to transform $x$ to $y$ using Bayesian model selection \cite{Peixoto2018PhysRevE}.
In this way, this stochastic block model can handle negative correlation values. With this method, one can determine community structure (i.e., blocks) including its number and hierarchical structure. 

\subsection{Partial correlation\label{sub:partial-corr}}

A natural method with which to avoid false positives due to indirect interaction effects in the Pearson correlation matrix is to use the partial correlation coefficient (as in, e.g., \cite{Delafuente2004Bioinfo,Salvador2005CerebCortex,Marrelec2006Neuroimage}). This entails measuring the linear correlation between nodes $i$ and $j$ after partialing out the effect of the other $N-2$ nodes. Specifically, to calculate the partial correlation between nodes $i$ and $j$, we first compute the linear regression of $X_i$ on $\{X_1, \ldots, X_N\}\setminus \{X_i, X_j\}$, which we write as $X_i \approx \sum_{m=1; m\neq i, j}^N \beta_{i,m} X_m$, where $\beta_{i,m}$ is the coefficient of linear regression. Similarly, we regress $X_j$ on $\{X_1, \ldots, X_N\}\setminus \{X_i, X_j\}$, which we write as $X_j \approx \sum_{m=1; m\neq i, j}^N \beta_{j,m} X_m$. The residuals for $L$ samples are given by $\varepsilon_{i,\ell} = x_{i\ell} - \sum_{m=1; m\neq i, j}^N \beta_{i,m} x_{m\ell}$ and $\varepsilon_{j,\ell} = x_{j\ell} - \sum_{m=1; m\neq i, j}^N \beta_{j,m} x_{m\ell}$, where $\ell \in \{1, \ldots, L\}$. The partial correlation coefficient, denoted by $\overline{\rho}_{ij}^{\text{par}}$, is the Pearson correlation coefficient between $\{\varepsilon_{i,1}, \ldots, \varepsilon_{i,L} \}$ and $\{\varepsilon_{j,1}, \ldots, \varepsilon_{j,L} \}$.

%
%
In fact, the partial correlation coefficient between $i$ and $j$ ($j\neq i$) is given by
\begin{equation}
\overline{\rho}_{ij}^{\text{par}} = - \frac{\Omega_{ij}} {\sqrt{\Omega_{ii} \Omega_{jj}}},
\label{eq:partial-corr-from-precision-matrix}
\end{equation}
where $\Omega = C^{-1}$ is the precision matrix~\cite{Whittaker1990book,Lauritzen1996book}. Equation~\eqref{eq:partial-corr-from-precision-matrix} implies that $\Omega_{ij}=0$ is equivalent to the lack of partial correlation, i.e., $\overline{\rho}_{ij}^{\text{par}} = 0$. This conditional independence property gives an interpretation of the precision matrix, $\Omega$.
Equation~\eqref{eq:partial-corr-from-precision-matrix} also implies that the partial correlation can be calculated only when $C$ is of full rank, whose necessary but not sufficient condition is $L \ge N$. If $C$ is rank-deficient, a natural estimator of 
the $N\times N$ partial correlation matrix $\overline{\rho}^{\text{par}} = (\overline{\rho}_{ij}^{\text{par}})$ is a pseudoinverse of $C$. However, the standard Moore-Penrose pseudoinverse is known to be a suboptimal estimator in terms of approximation error~\cite{Goldstein1974JRStatSocSerB,Hero2012IeeeTransInfoTh}, while the pseudoinverse is useful for screening for hubs in partial correlation networks~\cite{Hero2012IeeeTransInfoTh}.
If $C$ is of full rank, $\Omega$ as well as $C$ is positive definite. Therefore, although Eq.~\eqref{eq:partial-corr-from-precision-matrix} only holds true for $i\neq j$, if we denote the matrix defined by the right-hand side of Eq.~\eqref{eq:partial-corr-from-precision-matrix} including the diagonal entries by $\tilde{\rho}$, then the diagonal entries of $\tilde{\rho}$ are equal to $-1$, and $\tilde{\rho}$ is negative definite. We can verify this by rewriting Eq.~\eqref{eq:partial-corr-from-precision-matrix} as $\tilde{\rho} = - D^{-1/2} \Omega D^{-1/2}$, where
$D = \text{diag}(\Omega_{11}, \ldots, \Omega_{NN})$ is the diagonal matrix whose diagonal entries are $\Omega_{11}$, $\ldots$, $\Omega_{NN}$. 
If we consider matrix $\overline{\rho}^{\text{par}} = 2I + \tilde{\rho}$, where $I$ is the identity matrix, as a partial correlation matrix to force its diagonal entries to $1$ instead of $-1$, the eigenvalues of $\overline{\rho}^{\text{par}}$ are upper-bounded by $2$ \cite{Artner2022CommunStat}.

By thresholding the partial correlation matrix, or using an alternative method, one obtains an unweighted or weighted partial correlation network, depending on whether we further dichotomize the thresholded matrix. Because the partial correlation avoids the indirect interaction affect, the network created from random partial correlation matrices yields, for example, smaller clustering coefficients~\cite{Zalesky2012Neuroimage} than if we had used the Pearson correlation matrix.

While it apparently sounds reasonable to partial out the effect of the other nodes to determine a pairwise correlation between two nodes, it is not straightforward to determine when the partial correlation matrix is better than the Pearson correlation one. First, Eq.~\eqref{eq:partial-corr-from-precision-matrix} implies that extreme eigenvalues of $\overline{\rho}^{\text{par}}$ are those of a normalized precision matrix. Because the precision matrix is the inverse of the covariance matrix $C$, extreme eigenvalues of $\overline{\rho}^{\text{partial}}$ are derived from eigenvalues of $C$ with small magnitudes. It is empirically known for, e.g., financial time series, that small-magnitude eigenvalues of the covariance matrices are buried in noise, i.e., not distinguishable from eigenvalues of random matrices \cite{Laloux1999PhysRevLett,Plerou1999PhysRevLett}, as we discuss later in section~\ref{sub:null}. Therefore, the dominant eigenvalue of the precision matrix is strongly affected by noise \cite{Aste2006PhysicaA}. 

Second, the entries of $\overline{\rho}^{\text{par}}$ are more variable than those of Pearson correlation matrices. Specifically, if $(x_1, \ldots, x_N)$ obeys a multivariate normal distribution, the Fisher-transformed partial correlation of a sample partial correlation, i.e.,
\begin{equation}
z_{ij} = \frac{1}{2} \ln \left( \frac{1+\overline{\rho}_{ij}^{\text{par,sam}}}
{1-\overline{\rho}_{ij}^{\text{par,sam}}}
\right),
\end{equation}
where $\overline{\rho}_{ij}^{\text{par,sam}}$ is the sample 
partial correlation calculated through Eq.~\eqref{eq:partial-corr-from-precision-matrix} with $\Omega = \left(C^{\text{sam}} \right)^{-1}$, approximately obeys the normal distribution with
mean $\frac{1}{2} \ln
\left( \frac{1+\overline{\rho}_{ij}^{\text{par}}}
{1-\overline{\rho}_{ij}^{\text{par}}}
\right)$ and standard deviation
$[L-3-(N-2)]^{-1/2}$. This result dates back to Fisher (see e.g., \cite{Williams2022PsycholMethods}). In contrast, the corresponding result for the Fisher transformation of the Pearson correlation coefficient is that the transformed variable approximately obeys the normal distribution with mean $\frac{1}{2} \log
\left( \frac{1+\rho_{ij}^{\text{sam}}}
{1-\rho_{ij}^{\text{sam}}}
\right)$ and standard deviation
$(L-3)^{-1/2}$ 
\cite{Williams2022PsycholMethods}. Therefore, the partial correlation has more sampling variance than the Pearson correlation unless $L \gg N$.

Third, partial correlation matrices typically have more negative entries and smaller-magnitude entries than Pearson correlation matrices \cite{Millington2020ApplNetwSci,Williams2022PsycholMethods}. Combined with the larger variation of the sample partial correlation than the sample Pearson correlation discussed just above, the tendency that $\overline{\rho}^{\text{par}}_{ij}$ has a smaller magnitude than $\rho_{ij}$ poses a challenge of statistically validating the estimated partial correlation networks \cite{Williams2022PsycholMethods}.

Studies in neuroscience have compared partial correlation networks with simple correlation networks and/or with the corresponding underlying structural networks~\cite{Salvador2005CerebCortex,varoquaux2010brain,SmithMiller2011Neuroimage,Ryali2012Neuroimage,Brier2015Neuroimage,pervaiz2020optimising,liegeois2020revisiting} (see section~\ref{sec:brain}).
%
%
Some studies found that the similarity between partial correlation networks and structural networks is higher than that between Pearson correlation networks and structural networks~\cite{liegeois2020revisiting, Santucci2025biorxiv}.
%
%

\textit{Application example}: Wang, Xie, and Stanley analyzed correlation networks composed of stock market indices from 2005 to 2014 from 57 countries \cite{WangXieStanley2018ComputEcon}, widely covering the continents of the world, with each country corresponding to a node. They computed Pearson and partial correlation coefficients for the time series of the logarithmic returns, given by Eq.~\eqref{eq:log-return}, between each pair of countries, converted the correlation coefficient value into a distance (see Eq.~\eqref{eq:def-d_{ij}}), and constructed the minimal spanning trees, called the MST-Pearson and MST-Partial networks. These networks appeared to be scale-free (i.e., with a power-law-like degree distribution) trees. Among other things, they compared clusters and top centrality nodes (i.e., countries) between the MST-Pearson and MST-Partial. They observed that the results from the MST-Partial networks are more reasonable than those from the MST-Pearson construction in light of our general understanding of world economics.

\subsection{Graphical lasso and variants\label{sub:graphical-lasso}}

Estimating a true correlation matrix, which contains $N(N-1)/2$ unknowns, is an ill-founded problem unless the number of samples is sufficiently larger than $N(N-1)/2$. A strategy to overcome this problem is to impose sparsity of the estimated correlation network. A sparsity constraint enforces zeros on a majority of matrix entries to suppress the number of unknowns to be estimated relative to the number of samples.
%
Imposing sparsity on estimated correlation networks is a major form of covariance selection.
\emph{Structural learning} refers to estimation of an unknown network from data and usually assumes that the given data obey a multivariate normal distribution and that the estimated network is sparse. For reviews with tutorials and examples on this topic, see \cite{Drton2017AnnuRevStatAppl, Epskamp2018BehavRes, Epskamp2018PsycholMethods}.

The Gaussian graphical model assumes that 
the precision matrix from the data obeys a multivariate normal distribution and usually imposes sparsity of the precision matrix \cite{Lauritzen1996book}. In addition to reducing the number of unknowns to be estimated, a motivation behind estimating a sparse precision matrix is that $\Omega_{ij} = 0$ is equivalent to the absence of conditional linear dependence of the signals at the $i$th and $j$th nodes given all the other $N-2$ variables, which is easy to interpret.
The graphical lasso is an algorithm for learning the structure of a Gaussian graphical model~\cite{YuanLin2007Biometrika,Banerjee2008JMachineLearningRes,Friedman2008Biostatistics,FanFengWu2009AnnApplStat,Foygel2010Nips,Hastie2015book}. The graphical lasso maximizes the likelihood of the multivariate normal distribution under a lasso penalty (i.e., $\ell_1$ penalty), whose simplest version is of the form $\lambda \sum_{i,j=1}^N \left| \Omega_{ij} \right|$, where we recall that $\Omega_{ij}$ is the $(i, j)$ entry of the precision matrix, and $\lambda$ is a positive constant. This penalty term is added to the negative log likelihood to be minimized. If $\lambda$ is large, it strongly penalizes positive $\left| \Omega_{ij} \right|$, and the minimization of the objective function yields many zeros of the estimated $\Omega_{ij}$. The $(i, j)$ pairs for which the estimated $\Omega_{ij}$ is nonzero form edges of the network; if there is no edge between $i$ and $j$, they are conditionally independent, i.e., conditioned on the other $N-2$ variables. This conditional independence is also referred to as the pairwise Markov property because the distribution of $X_i$ only depends on $\{ X_j : (i, j) \text{ is an edge of the network}\}$. The pairwise Markov property is a special case of the global Markov property. The global Markov property dictates that $\{X_i : i \in A \}$ and $\{ X_i : i \in B \}$ are conditionally independent given $\{X_i : i \in S \}$ if $S$ is a cutset of $A$ and $B$, that is, any path in the graph connecting a node in $A$ and a node in $B$ passes through $S$. The pairwise and global Markov properties are major cases of the Markov random field, which is defined by a set of random variables having Markov property specified by an undirected graph \cite{Kindermann1980book, Lauritzen1996book, Rue2005book, Li2009book-Markov, WangKomodakis2013ComputVisImageUnderstanding}. The network is sparse by design. One can extend the lasso penalty function in multiple ways, for example, by allowing $\lambda$ to depend on $(i, j)$ and automatically determining $\lambda$ using an information criterion. Other ways to regularize the number of nonzero elements in the precision matrix than lasso penalty are also possible. (See e.g., \cite{Schafer2005Bioinfo,LiGui2006Biostat,Daspremont2008SiamJMatrixAnalAppl}.)

A neighborhood selection method is another algorithm to estimate a sparse Gaussian graphical model \cite{Meinshausen2006AnnStat}. With this method, one first carries out lasso regression for each $i$th variable (i.e., node) to determine a tentative small set of $i$'s neighbors. Second, if $j$ is a tentative neighbor of $i$, and $i$ is a tentative neighbor of $j$, then they are connected by an undirected edge $(i, j)$. The method named ``space'' (Sparse PArtial Correlation Estimation) advances the aforementioned neighborhood selection method by proposing to minimize a single composite penalized loss function, not separately minimizing the penalized loss function for each variable \cite{Peng2009JAmStatAssoc}. The loss function of ``space'' is a weighted sum of the regression error (i.e., error in estimating $X_i$ in terms of a linear combination of other $X_j$'s) over all $i$'s plus the usual $\ell_1$ penalty term. These regression-based methods as well the graphical lasso algorithms permit the case in which the number of observations, $L$, is smaller than the number of variables, $N$.

Bayesian variants of graphical lasso provide the level of undertainty in the estimated model. One such Bayesian approach assumes that we know node pairs that are not adjacent to each other, which is equivalent to imposing $\Omega_{ij} = 0$ for a given set of node pairs $(i, j)$ \cite{Hinne2014Neuroimage}. Such a situation is possible when both the correlation matrix and structural network are available, as is common in MRI experiments in the brain. The Bayesian method uses the G-Wishart distribution, which is the Wishart distribution constrained by $\Omega_{ij} = 0$ for non-adjacent node pairs in the given graph (i.e., structural network) $G$, as the prior distribution of the precision matrix, $\Omega$ \cite{Ataykayis2005Biometrika, Dobra2011JAmStatAssoc}. If $G$ is the complete graph, then the G-Wishart distribution is the Wishart distribution. Then, it uses data $(x_{i\ell}) \in \mathbb{R}^{N\times L}$ to update the distribution using Bayes' rule, obtaining the posterior distribution of $\Omega$, which is again a G-Wishart distribution but with updated parameter values. In other words, the G-Wishart distribution is a conjugate prior, giving us good intuitive pictures of how the prior distribution is transformed to the posterior distribution and mitigating the computational burden of the Bayesian method when $N$ is not small.

Gaussian graphical models assume normality. One approach to relax this assumption is to use a so-called nonparanormal distribution \cite{LiuLafferty2009JMachineLearningRes}. The idea is to assume that the transformed random variables $\left(f_1(X_1), \ldots, f_N(X_N)\right)$ obey a multivariate normal distribution, where $f_1$, $\ldots$, $f_N$ are monotone and differentiable functions. In this case, we say that the original variables $(X_1, \ldots, X_N)$ have a nonparanormal distribution or a Gaussian copula (given by $f_1$, $\ldots$, $f_N$, and the mean vector and the covariance matrix of the normal distribution after the transformation). The nonparanormal distribution is nonidentifiable. For example, if we replace $f_1(X_1)$ by a new function $\tilde{f}_1(X_1) \equiv 2 f_1(X_1)$ and appropriately scale the first row and column of the covariance matrix and the first entry of mean vector used for the multivariate normal distribution that the transformed $N$ variables obey, then one gets the same distribution of $(X_1, \ldots, X_N)$. Therefore, we further impose that the transformations $f_1$, $\ldots$, $f_N$ conserve the mean and variance of the original $X_1$, $\ldots$, $X_N$. Then, as in the case of Gaussian graphical models, $\Omega_{ij} = 0$ implies that $X_i$ and $X_j$ are conditionally independent of each other given all the other $N-2$ variables, in addition to that $f(X_i)$ and $f(X_j)$ are conditionally independent of each other. There are methods to estimate from data the functions $f_1$, $\ldots$, $f_N$ as well as a sparse covariance matrix $\Omega$, assuming a lasso penalty. Naturally, the nonparanormal distribution outperforms the graphical lasso when the true distribution of the data is not multivaraite normal \cite{LiuLafferty2009JMachineLearningRes}. Later work proposed algorithms for estimating the nonparanormal distribution with optimal convergence rate~\cite{LiuHanYuan2012AnnStat, XueZou2012AnnStat}; the idea behind the improved algorithms is to avoid explicitly calculating $f_1$, $\ldots$, $f_N$ and to exploit statistics of rank correlation coefficient estimators. There are also other methods for estimating non-normal graphical models without relying on copula
\cite{Morrison2017Nips, Baptista2024JMachineLearningRes}.

Extreme cases of non-Gaussian distribution of the random variables are discrete multivariate distributions, in particular when each $X_i$ is binary (i.e., $\in \{-1, 1 \}$.). In this case, the form of the distribution of ($X_1$, $\ldots$, $X_N$) proportional to $e^{\bm{x} \Omega \bm{x}^{\top}}$, where $\bm{x} = (X_1-\mu_1, \ldots, X_N - \mu_N)$, defines the Ising model. Note that the same form of distribution for continuous variables is the multivariate normal distribution, which we discussed above for graphical lasso. There are many methods for inferring the Ising model from observed multivariate binary data, which is the task referred to as the inverse Ising model and Boltzmann machine learning
\cite{Mora2011JStatPhys, Stein2015PlosComputBiol, Nguyen2017AdvPhys, Carleo2019RevModPhys, Mehta2019PhysRep}. Although exact likelihood maximization is available, it is practical only for small $N$. Therefore, various methods aim to overcome this limitation by allowing some approximation. Similar to graphical models, inference algorithms for the Ising model respecting the sparsity of the underlying network have also been investigated. For example, $\ell_1$-regularized methods based on pseudo-likelihood maximization were developed for estimating a sparse Ising Markov random field, or a graph in which the absence of the edge signifies the conditional independence between two nodes \cite{Bento2009Nips, Ravikumar2010AnnStat, Aurell2012PhysRevLett}. Decimation, or to recursively set the small edge weights to zero, combined with a stopping criterion and other heuristics, improves the performance of psuedo-likelihood maximization \cite{Decelle2014PhysRevLett}.

An alternative to the graphical lasso is to estimate sparse covariance matrices rather than sparse precision matrices under lasso penalty \cite{HuangLiu2006Biometrika,Bien2011Biometrika, LiuWangZhao2014JComputGraphicalStat, Wang2014StatComput,Wang2015BayesAnal,Kojaku2019ProcRSocA}. With this approach, the consequence of imposing sparsity, $C_{ij} = 0$, corresponds to marginal independence between $X_i$ and $X_j$. Similar to the case of the graphical lasso, one regards $(i, j)$ pairs for which the estimated $C_{ij}$ is nonzero as edges of the network.

Most graphical lasso models and their variants do not model the estimation problem relative to a null model correlation matrix. However, by estimating a sparse correlation matrix that is different relative to a null model of correlation matrix (see section~\ref{sub:null} for null models), it was found that the estimated correlation matrix gives a better description of the given financial correlation matrix data than the graphical lasso and that the choice of the null model also affects the performance \cite{Kojaku2019ProcRSocA}. By construction, this method infers a set of edges that are not expected from the so-called correlation matrix configuration model (see section~\ref{sub:null} for details).

\subsection{Statistical significance of correlation edges}

A test of significance of an edge, run on each edge, may sound like a natural way to filter a network. However, this idea is not easily feasible because multiple comparisons with $N(N-1)/2$ estimates, each of whose significance would have to be tested, is not practical given that $N(N-1)/2$ is usually large \cite{Epskamp2018BehavRes}. If we require a significance level of $0.05$, then Bonferroni correction for multiple comparisons implies that the edge statistic has to be tested to be significant with the $p$ value less than $0.05/[N(N-1)/2]$ for the edge to be actually significant at the $0.05$ significance level. Unless $N$ is small, this condition is usually too harsh. Furthermore, the different edges are correlated with each other, particularly if they share a node. In contrast, the Bonferroni correction assumes that the different tests are independent. Extensions of the Bonferroni corrections, such as the \v{S}id\'{a}k and Holm corrections, do not resolve these issues.

The false discovery rate approach, more specifically, the Benjamini-Yekutieli procedure \cite{Benjamini2001AnnStat}, provides a solution to these problems. This test is less stringent than family-wise error rate controls including the Bonferroni correction and allows dependency between different tests. To run this procedure to generate a correlation network, we first calculate the $p$ value for each node pair $(i, j)$. 
Second, we arrange all the $p$ values in ascending order, which we denote by $p_{(1)}$, $p_{(2)}$, $\ldots$, $p_{(N(N-1)/2)}$. Then, we reject the null hypothesis (i.e., regard that the edge with the corresponding $p$ value is significant) for $p_{(1)}$, $\ldots$, $p_{(M)}$, where
\begin{equation}
M = \max \left\{ i : p_{(i)} < \frac{0.05 i}{\frac{N(N-1)}{2} \sum_{i'=1}^{N(N-1)/2} \frac{1}{i'}} \right\}.
\label{eq:Benjamini}
\end{equation}
This procedure is a thresholding method with an automatically set threshold value implied by the number of edges, $M$, determined by Eq.~\eqref{eq:Benjamini}.
We note that the type of correlation matrix (e.g., Pearson or partial correlation) does not matter once $p$ values have been obtained. 

In the above procedure, what does it mean to calculate a $p$ value for a node pair? The answer varies. If we are given just one correlation matrix, which we assume in most of this article, the $p$ value can be that of the Pearson correlation coefficient in a standard textbook, if the Pearson correlation matrix is given as input \cite{Bassett2011PNAS} (see \cite{Achard2006JNeurosci, KangBowman2016Neuroimage, MaWangYe2020Microbiome} for different calculations of the $p$ value when a single matrix is given as input). If we are given multiple correlation matrices forming one group, the one-sample $t$-test provides a $p$ value for each $(i, j)$ pair, quantifying whether the $(i, j)$ correlation for the group is different from $0$ \cite{Salvador2005CerebCortex}. If correlation matrices from two groups need to be compared, the two-sample $t$-test provides a $p$ value for each $(i, j)$. In this case, the generated network will be a difference network, in which the edges represent node pairs for which the correlation is significantly different between the two groups.

In neuroimaging research, the network-based statistic (NBS) is popularly used for controlling for the same multiple comparison problem \cite{Zalesky2010Neuroimage}. In the NBS, one first calculates the $p$ value for each $(i, j)$, as in the Benjamini-Yekutieli procedure. Then, we keep $(i, j)$ whose $p$ values are smaller than an arbitrary threshold. If we just use those $(i, j)$ pairs to form a network, we would suffer from the aforementioned problems (i.e., multiple comparisons and dependencies between edges). Instead, the NBS focuses on the connected components induced by the surviving edges, which are small in number in general, and tests the significance of the sizes (i.e., the number of nodes) of the connected components using a permutation test. The NBS has been extended to a threshold-free method \cite{Baggio2018HumanBrainMapping}. NBS and many of its extensions are applicable to correlation matrix data beyond neuroscience. Note that the NBS is not a method to estimate a correlation network; it tactically avoids network estimation and the problem of multiple comparisons, while providing a statistically controlled downstream network analysis (i.e., test on the size of connected components). In this sense, the NBS casts a key question: why do we need to estimate a network first of all? We will discuss this topic in section~\ref{sub:recommendations}.

Covariance selection methods, such as the graphical lasso, do not explicitly test the significance of individual edges. The edges that survive these types of filtering methods should be regarded to be sufficiently strong to be included in the model \cite{Epskamp2018BehavRes}. 
	
\subsection{Temporal correlation networks\label{sub:temporal}}

Many empirical networks vary over time, including temporal correlation networks \cite{Masuda2020book}, and many methods have been developed for analyzing time-varying network data~\cite{HolmeSaramaki2012PhysRep,Holme2015EurPhysJB,Bazzi2016MultModelSimul,Masuda2020book}. If the given data is a multivariate time series that is non-stationary, then correlation matrices computed from the first 10\% of the time points may be drastically different from that computed from the last 10\%. So, there is the possibility of greater adaptability and better generalizability when one uses a time series of correlation networks rather than just one. One can then apply various temporal network analysis tools to the obtained temporal correlation networks.

A simple method to create dynamic correlation networks from multivariate time series data is sliding-window correlation \cite{Hindriks2016Neuroimage} (also called rolling-window correlation in the finance literature; see e.g. \cite{Adams2017JBankingFinance}). With this method, one considers time windows within the entire observation time horizon, $t=\{1, \ldots, t_{\max} \}$. These time windows may be overlapping or non-overlapping. Then, within each time window, one calculates the correlation matrix and network. If there are $100$ time windows within $[1, t_{\max}]$, then this method creates a temporal network of length $100$. Reliably computing a single correlation matrix and a static correlation network from multivariate time series requires a reasonable length (i.e., the number of time points) of a time window. Generation of a reliable dynamic correlation network requires longer data because one needs a multitude of such reasonably long time windows. A limitation of sliding-window correlations is that they are susceptible to large variability if the size of the time window is small, whereas a large window size sacrifices sensitivity to temporal changes~\cite{Lindquist2014Neuroimage}. 

Early seminal reports analyzed temporal correlation networks of stock markets by tracking financial indices and central nodes of the static correlation network over more than a decade~\cite{Onnela2002EurPhysJB,Onnela2003PhysRevE,Onnela2004EurPhysJB}. However, methods of dynamic correlation networks have been particularly developed in brain network analysis. In neuroimaging studies, in particular in fMRI studies, dynamic correlation networks are known as dynamic (or time-varying) functional connectivity networks \cite{Hutchison2013Neuroimage,Calhoun2014Neuron,Filippi2019FrontNeurosci,Hindriks2016Neuroimage,Lurie2020NetwNeurosci}. Temporal changes in functional (i.e., correlational) brain networks may represent neural signaling, behavioral performance, or changes in the cognitive state, for example. Patterns of time-varying functional networks may alter under a disease. One can also analyze stability of community structure of temporal correlation networks over time \cite{Bassett2011PNAS,Bassett2013PlosComputBiol,BraunSchafer2015PNAS}. (See also \cite{Bazzi2016MultModelSimul,Macmahon2015PhysRevX,almog2015mesoscopic,anagnostou2021uncovering,zema2021mesoscopic} for temporal stability analyses of community detection in financial correlation networks.) Many variations of the method, such as on how to create sliding windows and how to cluster time windows to define discrete-state transition dynamics, and change-point detection
%
%
are available in the field (e.g., see \cite{Lurie2020NetwNeurosci}). Many of these methods should be applicable to multivariate time series data to generate temporal correlation networks in other domains.

There are also methods for estimating dynamic precision matrices, proposed outside neuroscience \cite{Nakajima2013JBusiEconStat,Nakajima2015DigitalSignalProc,Hallac2017AcmSigkdd-lasso}. For example, the time-varying graphical lasso (TVGL) formulates the inference of discrete-time time-varying precision matrix as a convex optimization problem \cite{Hallac2017AcmSigkdd-lasso}. The objective function is composed of maximization of the log likelihood under a lasso sparsity constraint and the constraint that the precision matrix at adjacent time points does not change too drastically, enforcing the temporal consistency.

\section{Null models and random matrices\label{sub:null}}

In the analysis of (correlation) networks, a good practice to verify any structural or dynamical measurement $\alpha$ is to compare the value of $\alpha$ in the given network with the value of $\alpha$ achieved in random networks. This allows one to determine whether the $\alpha$ value measured for the given network data is explainable purely by the gross structural properties (e.g. edge density, degree distributions) of the random graph family (when the $\alpha$ value is similar between the given and random networks) or it is the result of other distinctive features of the data (when the $\alpha$ value is statistically different between the given and random networks). Many discoveries in network science owe to the fact that key analyses have prudently implemented this practice by using or inventing appropriate null models of networks. Already in one of the earliest seminal papers on small-world networks, Watts and Strogatz compared the average path length and the clustering coefficient of empirical networks with those of the Erd\H{o}s-R\'{e}nyi random graph having the same number of nodes and edges~\cite{Watts1998Nature}.

In this section, we report on similar concepts and results for correlation matrices. The idea is to introduce various null hypotheses that would imply certain properties for the sample correlation matrix, and compare the empirical matrix with the null model in order to extract statistically significant properties. This discussion will lead us to consider on one hand models defined in analogy with null models for networks, and on the other hand genuine models for correlation matrices derived from RMT.
Several papers have noted the need for proper null models specifically for correlation networks \cite{Faust2012NatRevMicrobiol,Zalesky2012Neuroimage,Macmahon2015PhysRevX,Masuda2018PhysRevE,Vasa2022NatRevNeurosci}. We should use correlation networks derived from a random correlation matrix as a null model. We stress that a random correlation matrix is different from a random network model (e.g., Erd\H{o}s-R\'{e}nyi model), because of the dependencies between entries. Similarly, many classes of random matrices are not appropriate null models for correlation networks, either. For example, a symmetric matrix whose all on-diagonal entries are $1$ and off-diagonal entries are i.i.d.\,uniformly on $(-1, 1)$ is almost never a correlation matrix unless $N$ is small \cite{Bohm2014StatProbLett}. In this section, we introduce several null models of correlation matrices. 
All the null models presented give distributions over correlation matrices. Then, using any method introduced in previous sections (e.g., by thresholding in various ways), one can define corresponding null models over correlation networks.

\subsection{Models based on shuffling\label{sec:homogeneous}}

A straightforward and traditional null model consists in shuffling the original data, $\{ x_{i\ell} \}$, based on which the correlation matrix is calculated. This method is especially typical for multivariate time series data. In the simplest case, one randomizes all entries independently within each time series, thereby destroying all the cross-correlations while preserving the original values for each time series separately.

As a more constrained option, one preserves the power spectrum of the time series at each node while the time series is otherwise randomized \cite{Theiler1992PhysicaD,Schreiber2000PhysicaD,Zalesky2012Neuroimage,Vasa2022NatRevNeurosci}. More specifically, one Fourier transforms the time series at the $i$th node, randomize the phase, and carry out the inverse Fourier transform. Then, for the randomized multivariate time series, one calculates the correlation matrix, which is used as control.

Another method that preserves the full autocorrelation structure within each single time series, while randomizing cross-correlations among the $N$ time series, has been proposed in~\cite{iyetomi2011causes,iyetomi2011fluctuation}. The method is called the rotational random shuffling (RRS) model because it first imposes periodic boundary conditions (i.e., `gluing' the last timestep to the first one) to turn each time series into a `ring', and then randomly rotates the $N$ rings with respect to each other while keeping each ring internally intact.
%

One can impose additional constraints on the randomization of time series depending on the properties other than the power spectrum that one wants to have the null model preserve \cite{Shinn2023NatNeurosci}.

\subsection{Models inspired by network analysis\label{sec:configuration}}

Other null models are explicitly inspired by null models that are routinely used for networks. For instance, the H--Q--S algorithm, invented by Hirschberger, Qi, and Steuer \cite{Hirschberger2007EurJOperRes} is an equivalent of the Erd\H{o}s-R\'{e}nyi random graph in general network analysis. Specifically,
given the covariance matrix, $C^{\text{sam}}$, the H--Q--S algorithm generates random covariance matrices, $C^{\rm HQS}$, under the following constraints.
First, the expectation of each on-diagonal entry of $C^{\rm HQS}$ is equal to the average of the $N$ on-diagonal entries of $C^{\text{sam}}$, denoted by $\mu_{\rm on} \equiv (1/N) \times \sum_{i=1}^N C^{\text{sam}}_{ii}$.
Second, the expectation and variance of each off-diagonal entry of $C^{\rm HQS}$ are equal to those of $C^{\text{sam}}$ 
calculated on the basis of all the off-diagonal entries, denoted by $\mu_{\rm off}$ and $\sigma_{\rm off}^2$, respectively.
Optionally, one can also constrain the variance of the on-diagonal entries of $C^{\rm HQS}$ \cite{Hirschberger2007EurJOperRes} or
use a fine-tuned heuristic variant of the algorithm \cite{Zalesky2012Neuroimage}. To implement the most basic H--Q--S algorithm without constraining the variance of the on-diagonal entries of  $C^{\rm HQS}$, we set
\begin{equation}
L^{\rm HQS} \equiv \max \left( 2, \lfloor \left(\mu_{\rm on}^2 - \mu_{\rm off}^2\right)/\sigma_{\rm off}^2\rfloor\right),
\end{equation}
where $\lfloor \cdot \rfloor$ is the largest integer that is not greater than the argument.
Then, we draw $N \times L^{\rm HQS}$ variables, denoted by $x_{i\ell}$ (with $i \in \{1, \ldots, N\}$ and $\ell \in \{ 1, \ldots, L^{\rm HQS}\}$), independently from the identical normal distribution with mean $\sqrt{\mu_{\rm off} / L^{\rm HQS}}$ and variance $-\mu_{\rm off} / L^{\rm HQS} +
\sqrt{\mu_{\rm off}^2 / (L^{\rm HQS})^2 + \sigma_{\rm off}^2 / L^{\rm HQS}}$. Then, the H--Q--S algorithm sets the covariance matrix by
\begin{equation}
C^{\rm HQS}_{ij} = \sum_{\ell=1}^{L^{\rm HQS}} x_{i\ell} x_{j\ell}\quad
i, j \in \{1, \ldots, N\}.
\end{equation}
It is known that $\langle C^{\rm HQS} \rangle_{ij} = \delta_{ij} \mu_{\rm on} + (1-\delta_{ij}) \mu_{\rm off}$. Therefore, 
%
%
the expectation of the correlation matrix, $\rho^{\text{HQS}}$, is approximately given by $\langle\rho^{\rm HQS}\rangle_{ij} = \delta_{ij} + (1-\delta_{ij}) \mu_{\rm off}/\mu_{\rm on}$.
This highlights the completely homogeneous nature of the null model.
For instance, while the degree of empirical correlation networks is usually heterogeneously distributed~\cite{Bonanno2003PhysRevE}, this property is not captured by the H--Q--S algorithm \cite{Fornito2013Neuroimage}. 

One of the most popular heterogeneous null models for networks is the configuration model, i.e., a uniform ensemble of networks under the constraint that the degree of each node is conserved~ \cite{Fosdick2018SiamRev,Vasa2022NatRevNeurosci}, either exactly or in expectation. By comparing given networks against a configuration model, one can reliably quantify and discuss various network properties such as network motifs \cite{Milo2002Science}, community structure \cite{Fortunato2010PhysRep}, rich clubs \cite{Colizza2006NatPhys}, and core-periphery structure \cite{Kojaku2018NewJPhys}. The rationale behind the use of the configuration model is that the node’s degree is heterogeneously distributed for many empirical networks and that one generally wants to explore structural, dynamical, or other properties of networks that are not immediate outcomes of the heterogeneous degree distribution. One can extend the configuration model by imposing additional constraints that the network under discussion is supposed to satisfy, such as spatiality, i.e., the constraint that the nodes are embedded in a metric space and the probability of an edge depends on the distance between the two nodes \cite{Expert2011PNAS}. See \cite{Squartini2015NewJPhys,Squartini2017book,Fosdick2018SiamRev} for reviews of configuration models and best practices for generating random realizations from such models.

We should similarly test properties found for given correlation networks against appropriate null models. However, the usual configuration models are not appropriate as null models of correlation networks because they are significantly different from correlation networks derived from purely random data \cite{Steuer2006BriefBioinfo,Zalesky2012Neuroimage,Macmahon2015PhysRevX,Bazzi2016MultModelSimul}. The expectation $\langle A_{ij}\rangle$ of the $(i,j)$ entry of the adjacency matrix of the configuration model conditioned on the degrees of all nodes, at least in the idealized and unrealistic regime of weak heterogeneity of the degrees~\cite{Squartini2015NewJPhys,Squartini2017book}, is equal to 
\begin{equation}
    \langle A_{ij}\rangle=\frac{k_i k_j}{N\overline{k}},\label{eq:cm}
\end{equation}
%
%
where $k_i=\sum_{j=1}^N A_{ij}$ is the degree of the $i$th node in the original network, $A_{ij}$ is the entry of the \emph{empirical} adjacency matrix of the original network (i.e., $A_{ij}=1$ if there is an edge between the $i$th and $j$th nodes, and $A_{ij}=0$ otherwise), and $\overline{k}$ is the average degree over all nodes. The above expected value also represents the probability of independently connecting the $i$th and $j$th nodes in realizations of the configuration model for networks. 
Note that, one can indeed realize graphs in the configuration model by sampling edges independently (at least if the constraint on the degree is `soft', i.e. realized only as an ensemble average over realizations~\cite{Squartini2017book}). However, correlation matrices cannot be generated with independent entries, even in the null model of independent signals.
This is because, even under the null hypothesis of independent realizations of the original time series, the correlation matrix constructed from such time series still obeys the `metric' (or triangular) inequality in~\eqref{eq:Langford-bound}. We will elaborate more on this point below.

To see what Eq.~\eqref{eq:cm} yields by merely replacing an empirical adjacency matrix with an empirical correlation matrix $\rho_{ij}$, we proceed as follows~\cite{Macmahon2015PhysRevX}. 
We assume that each empirical signal is standardized in advance such that $\text{Var}(X_i)=1$, $\forall i\in \{1, \ldots, N\}$. In this way, we do not need to distinguish between the correlation (i.e., $\rho_{ij}$) and covariance (i.e., $C_{ij}$) matrices.
We express the `degree' as 
\begin{equation}
k_i = \sum_{j=1}^N \rho_{ij} = \sum_{j=1}^N C_{ij}
= \sum_{j=1}^N \text{Cov}(X_i, X_j) 
= \text{Cov}(X_i, X_{\text{tot}}),
\label{eq:Bazzi2016-ki}
\end{equation}
where $\text{Cov}$ represents the covariance and $X_{\text{tot}}=\sum_{i=1}^N X_i$ is the `total' signal.
Then, we obtain
\begin{equation}
N \overline{k} = \sum_{i=1}^N k_i = \text{Cov}(X_{\text{tot}}, X_{\text{tot}}) = \text{Var}(X_{\text{tot}}),
\label{eq:Bazzi2016-N<k>}
\end{equation}
where $\text{Var}(X_{\text{tot}})$ is the variance of $X_{\text{tot}}$. By inserting the above quantities into~\eqref{eq:cm} with $A_{ij}$ replaced by $\rho_{ij}$, we obtain for the expected correlation matrix \begin{equation}
\langle \rho_{ij}\rangle= \frac{\text{Cov}(X_i, X_{\text{tot}}) \text{Cov}(X_j, X_{\text{tot}})}
{\text{Var}(X_{\text{tot}})}
= \frac{\text{Cov}(X_i, X_{\text{tot}})}
{\sqrt{\text{Var}(X_{\text{tot}})} \sqrt{\text{Var}(X_i)}}
\cdot \frac{\text{Cov}(X_j, X_{\text{tot}})}
{\sqrt{\text{Var}(X_{\text{tot}})} \sqrt{\text{Var}(X_j)}}
= 
\rho(X_i, X_{\text{tot}}) \rho(X_j, X_{\text{tot}}).
\label{eq:Bazzi2016-config}
\end{equation}
Technically, this expected matrix is a covariance matrix because it is symmetric, rank 1, and the only nonzero eigenvalue is positive \cite{Macmahon2015PhysRevX,Valdano2019PhysRevX}. 
To interpret the meaning of the above expression for $\langle\rho_{ij}\rangle$, we recall the definition of the conditional three-way partial Pearson correlation coefficient~\cite{Whittaker1990book,Anderson2003book}:
\begin{equation}
\rho(X_i, X_j \mid X_{\text{tot}})
= \frac{\rho(X_i, X_j) - \rho(X_i, X_{\text{tot}}) \rho(X_j, X_{\text{tot}})}
{\sqrt{1-\rho(X_i, X_{\text{tot}})^2}
\sqrt{1-\rho(X_j, X_{\text{tot}})^2}
}.
\label{eq:three-way-partial-corr}
\end{equation}
We therefore conclude that the expected correlation matrix in~\eqref{eq:Bazzi2016-config} is a correlation matrix of $N$ signals that satisfy the conditional independence relationship 
\begin{equation}
\rho(X_i, X_j \mid X_{\text{tot}} ) = 0
\label{eq:Bazzi2016-independence}
\end{equation}
$\forall i, j (\neq i) \in \{1, \ldots, N \}$~\cite{fenn2012dynamical,Bazzi2016MultModelSimul}.

However, when one generates a correlation network from the configuration model, i.e. from a correlation matrix obeying~\eqref{eq:Bazzi2016-independence}, the generated network is far from a typical correlation network generated by random data, due to the triangular inequality mentioned above. To see an example of this, let us revisit the example briefly explained in section~\ref{sub:dichotomizing}. Let us consider purely random data in which we generate each sample $x_{i\ell}$, $i \in \{1, \ldots, N \}$, $\ell \in \{1, \ldots, L\}$ (where we recall that $L$ is the number of samples for each node) as i.i.d.\,random numbers obeying a given distribution. Then, we calculate the sample correlation matrix for $\{x_{i\ell} \}$ and then a sample correlation network. This procedure immediately establishes a connection between null models for sample correlation matrices and RMT~\cite{mehta2004random,livan2018introduction,potters2020first}, some elements of which are discussed in section~\ref{sec:RMT}. For a broad class of methods, including dichotomizing, the generated correlation network has a high clustering coefficient~ \cite{Zalesky2012Neuroimage}, precisely because of the inequality~\eqref{eq:Langford-bound}. Therefore, high clustering coefficients in correlation networks should not come as a surprise. In contrast, networks generated by the ordinary configuration model yield low clustering coefficients \cite{Newman2018book}, disqualifying it as a null model for correlation networks~\cite{Macmahon2015PhysRevX}. 
If we use the usual configuration model as the null model, we would incorrectly conclude that a given correlation network has high clustering even if the network does not have particularly high clustering among correlation networks. 
The configuration model as null model also underperforms 
the simpler null model, called the uniform null model (which is analogous to the random regular graph and to the H--Q--S algorithm explained above) on benchmark problems of community detection when communities of different sizes are present in single networks and those communities are detected by modularity maximization~\cite{Bazzi2016MultModelSimul}. 

A different configuration model, specifically designed for covariance matrices, can be defined as follows. This model preserves the expectation of each row sum excluding the diagonal entry, which is equivalent to each node's degree in the case of the adjacency matrix of a conventional network \cite{Masuda2018PhysRevE}. This algorithm, which we refer to as the correlation matrix configuration model, preserves the expectation of each diagonal entry of $C^{\text{sam}}$, or the variance of each variable, and the expectation of each row sum excluding the diagonal entry, i.e., $\sum_{j=1; j\neq i}^N C^{\text{sam}}_{ij}$ $\forall i$, corresponding to the degree of the $i$th node. Under these constraints, the correlation matrix configuration model uses the distribution of $x_{i\ell}$, determined from the maximum entropy principle. In fact, each $(x_{1\ell}, \ldots, x_{N\ell})^{\top}$, $\ell \in \{1, \ldots, L\}$, independently obeys an identical multivariate normal distribution whose mean is the zero matrix and covariance matrix is denoted by $C^{\text{cm}}$. Therefore, the correlation matrix configuration model is the Wishart distribution $W_N(C, L)$ that we have introduced in section~\ref{sub:estimation-covariance-matrix}, with $C=C^{\text{cm}}$. The matrix $C^{\text{cm}}$ is of the form
$\left[(C^{\text{cm}})^{-1}\right]_{ij} = -\left[ \delta_{ij} \cdot 2 \alpha_i  + (1-\delta_{ij})(\beta_i + \beta_j) \right]$, where $\delta_{ij}$ is the Kronecker delta symbol, and $\alpha_i$ and $\beta_i$ are parameters to be determined.
One determines the values of $\alpha_i$ and $\beta_i$ using a gradient descent algorithm \cite{Masuda2018PhysRevE} or by reformulating the problem as a convex optimization problem and solving it~\cite{Kojaku2019ProcRSocA}.

\subsection{Models based on random matrix theory\label{sec:RMT}}

Another class of null models is based on powerful estimates provided by RMT for the expected spectral properties of a random sample correlation matrix, rather than for the expected matrix itself~\cite{mehta2004random,livan2018introduction,potters2020first}. Note that the expected correlation matrix, under the null hypothesis of $N$ independent signals, is the identity matrix whose entries we denote as  $\rho^{\text{MG1}}_{ij}=\delta_{ij}$~\cite{Macmahon2015PhysRevX,Bazzi2016MultModelSimul}. Equivalently, $\rho^{\text{MG1}}=I$ where $I$ is the $N\times N$ identity matrix. This null model corresponds to an expectation under white noise signals $\{x_{i\ell}\}$ that are independent for different nodes $i \in \{1, \ldots, N\}$ and samples $\ell \in \{1, \ldots, L\}$. 
However, the sample pairwise correlation $\rho^{\text{sam}}_{ij}$ measured from the signals, even under the null hypothesis of independence, will be different from the identity matrix unless $L\to\infty$, assuming a finite $N$. To take into account the effects of finite $L$, it is convenient to look at the distribution of the random sample correlation matrix. If we assume a standardized independent normal distribution for $x_i$ $\forall i$, then $\rho^{\text{sam}}$ obeys the Wishart distribution with $C=I$, i.e. $W_N(I, L)$. Note that the expectation of $W_N(I, L)$ is $I$.
The ensemble of Wishart matrices is a well studied topic in RMT. It turns out that, in addition to the distribution $W_N(I, L)$ for the entire sample correlation matrix, one can accurately describe the limiting density of eigenvalues in the asymptotic limit $N\to\infty$ and $L\to\infty$, with $L/N\to Q>1$. Note that $Q>1$ is a necessary condition for the sample correlation matrix to be non-degenerate, as we already mentioned. The limiting eigenvalue density $p_Q(\lambda)$, known as the Marchenko-Pastur distribution, has the form 
\begin{equation} p_Q(\lambda)=
\begin{cases}
\frac{Q\sqrt{(\lambda_+-\lambda)(\lambda-\lambda_-)}}{2\pi\lambda}&\text{for } \lambda_-<\lambda<\lambda_+,\\
0&\text{otherwise},
\end{cases}
\label{eq:MP}
\end{equation}
where
\begin{equation}
\lambda_\pm = \left(1 \pm \sqrt{Q^{-1}} \right)^2
\label{eq:lambda-pm-Q}
\end{equation}
are the expected minimum ($\lambda_-$) and maximum ($\lambda_+$) eigenvalues.
As $Q\to \infty$, which corresponds to an infinite number of samples per node, it holds true that $\lambda_\pm\to 1$. This result implies that all eigenvalues become unity, and it is because each sample correlation matrix becomes the identity matrix in that case, as we already mentioned; then all eigenvalues are equal to 1. 
The fact that $Q$ is necessarily finite for empirical correlation matrices makes Eq.~\eqref{eq:MP} particularly useful as a null model for the empirical eigenvalue distribution expected under the hypothesis of independence of the $N$ observed time series. In particular, early studies of financial multivariate time series data found that only a small number of the largest eigenvalues of the empirical covariance matrix are found above $\lambda_+$~\cite{Laloux1999PhysRevLett,Plerou1999PhysRevLett}. In other words, only a few leading eigenvalues and the associated eigenvectors outside the prediction of RMT are statistically significant relative to the null model. 
As we discuss below, the role of the largest empirical eigenvalue is however different from that of the next largest ones. Specifically, the former encodes a system-wide, overall positive correlation, while the latter represent `mesoscopic' information arising from the presence of internally coherent \emph{groups} of time series. Based on this interpretation, RMT has provided useful null models for covariance/correlation matrix data, in particular in financial time series data analysis;
see \cite{Bun2017PhysRep} for a review. 
In neuroscience, RMT has been applied only more recently (e.g., for noise-cleaning in the estimation of correlation matrices~\cite{ibanez2023noise} as we describe later in this section and for community detection~\cite{almog2019uncovering} as we describe in section~\ref{sec:community}) and less systematically. Nonetheless, it holds promise as a general and powerful tool for the analysis of large data in virtually any field, especially in the modern era of data science~\cite{potters2020first}.
Also see \cite{Holmes1991SiamJMatAnalAppl} for a review of random correlation as opposed to covariance matrices. 

Among many possible specific choices of null models for correlation matrices based on RMT, here we consider two models, which we denote by $\rho^{\rm MG2}$ and $\rho^{\rm MG3}$, proposed in~\cite{Macmahon2015PhysRevX, almog2019uncovering}. A given correlation matrix is symmetric and positive semidefinite and therefore can be decomposed as
\begin{equation}
\rho^{\text{sam}} = \sum_{i=1}^N \lambda_i \bm u_{(i)} \bm u_{(i)}^{\top},
\label{eq:eigen-decomposition-MG}
\end{equation}
where $\lambda_i (\ge 0)$ is the $i$th eigenvalue of $\rho^{\text{sam}}$, and $\bm u_{(i)}$ is the associated normalized right eigenvector. The null model correlation matrix $\rho^{\rm MG2}$ preserves the contribution of small eigenvalues to Eq.~\eqref{eq:eigen-decomposition-MG}, which are regarded to be noisy and described by RMT, and is given by
\begin{equation}
\rho^{\rm MG2} = \sum_{i: \lambda_i \le \lambda_+} \lambda_i \bm u_{(i)} \bm u_{(i)}^{\top}.
\end{equation}
The boundary $\lambda_+$ originates from the Marchenko-Pastur distribution  given by Eq.~\eqref{eq:MP} and, as we mentioned, represents the expected largest eigenvalue \emph{under the null hypothesis of independent signals}. Matrix $\rho^{\rm MG2}$ is not a correlation matrix because its diagonal elements are not equal to 1. However, this does not affect most network analyses because we usually ignore diagonal elements or self-loops in correlation networks. Matrix $\rho^{\rm MG2}$ represents a null model constructed only from the eigenvalues of the empirical correlation matrix that are deemed to be noisy. Therefore, comparing the empirical matrix against $\rho^{\rm MG2}$ singles out properties that cannot be traced back to noise \cite{Macmahon2015PhysRevX}. Note that the difference 
\begin{equation}
    \rho^{\text{sam}}-\rho^{\rm MG2}= \sum_{i: \lambda_i > \lambda_+}\lambda_i \bm u_{(i)} \bm u_{(i)}^{\top}\label{eq:PCA}
\end{equation}
between the sample correlation matrix $\rho^{\text{sam}}$ and the null model $\rho^{\rm MG2}$ is nothing but the sum of the dominant eigencomponents of $\rho^{\text{sam}}$. This matrix coincides with the output of the popular PCA technique, also called the eigenvalue clipping. The main difference between the generic use of that technique and the one based on $\rho^{\rm MG2}$ in Eq.~\eqref{eq:PCA} is the criterion (here based on RMT) for the selection of the number of principal components to retain.

By contrast, the matrix $\rho^{\rm MG3}$ also preserves the contribution of the largest eigenvalue in addition to that of the noisy eigenvalues~\cite{Macmahon2015PhysRevX,almog2015mesoscopic,anagnostou2021uncovering,zema2021mesoscopic}. The largest eigenvalue is useful to control for separately, if all the entries of the associated eigenvector are positive. It is because, in that case, it represents the effect of a common trend in the original time series,
%
%
giving an uninformative all-positive overall contribution (also called global mode, or market mode in the context of financial time series) to the sample correlation matrix. The matrix $\rho^{\rm MG3}$ is given by
\begin{equation}
\rho^{\rm MG3} = \lambda_{\max} \bm u_{(\max)} \bm u_{(\max)}^{\top} +
\sum_{i: \lambda_i \le \tilde{\lambda}_+} \lambda_i \bm u_{(i)} \bm u_{(i)}^{\top},\label{eq:MG3}
\end{equation}
where $\max$ is the index for the dominant eigenvalue of $\rho^{\text{sam}}$, and ${\lambda}_+$ has been replaced by 
\begin{equation} \tilde{\lambda}_+\equiv \left( 1-\frac{\lambda_{\max}}{N} \right)~{\lambda}_+\label{eq:tilde}.
\end{equation}
The above modification in the value of the threshold eigenvalue, with respect to the previous null model in Eq.~\eqref{eq:lambda-pm-Q},
%
%
originates from the fact that sample correlation matrices always have trace equal to $N$, with all their diagonal entries being unity by construction. Therefore, the addition of a global mode represented by a large $\lambda_{\max}$ has the unavoidable effect of proportionally reducing all the other eigenvalues in such a way that the trace is preserved~\cite{Laloux1999PhysRevLett,almog2019uncovering}. The value in Eq.~\eqref{eq:tilde} thus represents the expected largest eigenvalue \emph{under the null hypothesis of independent signals plus a global mode}. Note that $\tilde{\lambda}_+<{\lambda}_+$, which implies that any empirical eigenvalue $\lambda_i$  smaller than $\lambda_+$ but  larger than $\tilde{\lambda}_+$, i.e. $\tilde{\lambda}_+<\lambda_i<\lambda_+$, is interpreted as noisy (hence discarded) under the null model $\rho^{\rm MG2}$ and as informative (hence retained) under the null model $\rho^{\rm MG3}$~\cite{almog2019uncovering}. Also note that all the entries of the dominant eigenvector, $\bm u_{(\max)}$, are positive, which is a necessary condition for the null model $\rho^{\rm MG3}$ to have a clear interpretation, if there is a sufficiently strong common trend affecting all the $N$ signals. If this common trend is so strong that all the entries of the correlation matrix are positive, then the Perron-Frobenius theorem ensures the positivity of the dominant eigenvector. 
When this happens, the common trend obscures all the mutual correlations among the signals. Matrix $\rho^{\rm MG3}$ deliberately removes this global trend in addition to the noise, to reveal the underlying structure.
Therefore, properties of correlation matrices or correlation networks that are not expected from $\rho^{\text{MG3}}$
%
%
represent those not anticipated by the 
simultaneous presence of local noise and global trends. In other words, they reflect the presence of mesoscopic correlated structures such as correlation-induced communities~\cite{Macmahon2015PhysRevX,almog2015mesoscopic,almog2019uncovering,anagnostou2021uncovering,mircea2022phiclust,zema2021mesoscopic}. As we will discuss in section~\ref{sec:community}, one can indeed use matrix $\rho^{\rm MG3}$ to successfully detect communities of correlated signals.

We finally discuss another RMT-based model that, rather than representing a null model of the data, aims at providing the best fit, i.e., an optimal estimate, for the true (unobserved) correlation matrix $\rho^{\text{true}}$, starting from the sample correlation matrix $\rho^{\text{sam}}$. The model is called the optimal rotationally invariant estimator (RIE) for the correlation matrix~\cite{Bun2017PhysRep,ibanez2023noise}. Informally, the RIE estimator is the matrix $\rho^{\text{RIE}}$ that, among the correlation matrices sharing the same eigenvectors with $\rho^{\text{sam}}$, achieves the minimum Hilbert-Schmidt distance $d_{\text{HS}}(\rho, \rho^{\text{true}}) = \text{Tr}
[(\rho -\rho^{\text{true}})^2]$ from the true population matrix $\rho^{\text{true}}$. This task might seem impossible at first sight, because $\rho^{\text{true}}$ is unobservable. However, it turns out that, for the minimization of $d_{\text{HS}}(\rho, \rho^{\text{true}})$ for large enough $L$, it is sufficient to know the spectral density $p^{\text{true}} (\lambda)$ of $\rho^{\text{true}}$, which can be obtained from the spectral density $p^{\text{sam}} (\lambda)$ of $\rho^{\text{sam}}$, thanks to a type of self-averaging property~\cite{Bun2017PhysRep}. 
The final ingredient consists in modifying the eigenvalues $\{\lambda_{i}\}_{i=1}^N$ of $\rho^{\text{sam}}$ to
\begin{equation}
\tilde{\lambda}_i\equiv \frac{\lambda_{i}}{|1-Q^{-1}+ Q^{-1}z_i s(z_i)|^2}\qquad i=1,N\label{eq:cauchy}
\end{equation} 
where $z_i\equiv \lambda_i-i\eta$ is the complexification of $\lambda_i$; $\eta$ is a small parameter that should vanish in the $N\to\infty$ limit; $s(z)\equiv \text{Tr}[(z I-\rho^{\text{sam}})^{-1}]/N$ is the so-called Cauchy transform of the spectrum of $\rho^{\text{sam}}$~\cite{Bun2017PhysRep}. A convenient choice for $\eta$ in the case of finite $N$ is $\eta=N^{-1/2}$~\cite{Bun2017PhysRep}. (However, see~\cite{ibanez2023noise} and the application example below for an improved variant.)
Using Eq.~\eqref{eq:cauchy}, we obtain the optimal RIE as
\begin{equation}
\rho^{\text{RIE}} = \sum_{i=1}^N \tilde{\lambda}_i \bm u_{(i)} \bm u_{(i)}^{\top}.
\label{eq:RIE}
\end{equation}
For large $L$, the optimal RIE is expected to outperform, in terms of $d_{\text{HS}}$, any other estimator (including PCA, shrinkage, and other corrected sample estimators) that, like the RIE itself, modifies only the spectrum of $\rho^{\text{sam}}$ and not its eigenvectors~\cite{Bun2017PhysRep}.
%

\emph{Application example.} Ib\`a\~nez-Berganza et al.~\cite{ibanez2023noise} compared several noise-cleaning methods of estimation of covariance and precision matrices from both human brain activity (fMRI) time series and synthetic data of size comparable with that typically encountered in neuroscience. They assessed the reliability of each method via the test-set likelihood and, in the case of synthetic data, via
%
%
the distance from the true precision matrix. The methods considered include the eigenvalue clipping (or PCA; see above), linear shrinkage (see section~\ref{sub:estimation-covariance-matrix}), graphical lasso (see section~\ref{sub:graphical-lasso}), FA~\cite{Anderson2003book,Mulaik2010book}, early-stopping gradient ascent algorithms, the RMT-based optimal RIE given by Eq.~\eqref{eq:RIE}, and a variant of the last one where the parameter $\eta$
is optimized by cross-validation on a grid of values
rather than being set to $\eta=N^{-1/2}$ (see above). Their cross-validated RIE outperformed all the other estimators in the severe undersampling regime (i.e., small $Q$) typical of fMRI time series, highlighting the power of RMT for the analysis of neuroscience data. Notably, the cross-validated RIE was the only method that improved upon the raw sample correlation matrix $\rho^{\text{sam}}$ in the estimation of the true correlation matrix $\rho^{\text{true}}$, in all the simulated regimes and especially for strongly correlated synthetic data. They finally proposed a simple algorithm based on an iterative likelihood gradient ascent, leading to accurate estimations in weakly correlated synthetic data sets. A Python code implementing all the methods used in the paper is available~\cite{Lucibello}.

\section{Network-analysis inspired analysis directly applied to correlation matrices\label{sub:inspired}}

As we already mentioned, a straightforward way to use null models for correlation matrices as controls of correlation networks is to generate correlation networks from the correlation matrix generated by the null model or its distribution. In this section, we showcase another usage of null models for correlation matrices, which is to conduct analysis inspired by network analysis but directly on correlation matrix data with the help of null model correlation matrices. Crucially, this scenario does not involve transformation of a given correlation matrix into a correlation network. To explain the idea, consider financial time series analysis using correlation matrices. Portfolio optimization and RMT directly applied to correlation matrix data are among powerful techniques to analyze such data~\cite{livan2018introduction,potters2020first,Bun2017PhysRep}. These methods do not suffer from difficulties in transforming correlation matrices into correlation networks because they do not carry out such a transformation. In contrast, a motivation behind carrying out such a transformation is that one can then use various network analysis methods. A strategy to take advantage of both approaches is to adapt network analysis methods for conventional networks to the case of correlation matrix data.

\subsection{Degree}

Many empirical networks show heterogeneous degree distributions such as a power-law-like distribution~\cite{Caldarelli2007book, Barabasi2016book, Newman2018book, Menczer2020book}; such networks are called scale-free networks. The same holds true for the weighted degree of many networks~\cite{Barrat2004PNAS}. Correlation networks are no exception, not much depending on how one constructs a network from correlation matrix data \cite{Onnela2003PhysRevE, Eguiluz2005PhysRevLett, ZhangHorvath2005StatApplGenetMolBiol, Bassett2008JNeurosci, Masuda2018PhysRevE}.

If we do not transform the given correlation matrix into a network, the node's weighted degree represents how the node's signal, $X_i$, is close to the signal averaged over all the nodes, $X_{\text{total}}$, as shown in Eq.~\eqref{eq:Bazzi2016-ki}. Previous research showed that the weighted degree calculated in this manner is heterogeneously distributed for some empirical data, while the right tail of the distribution is not as fat as typical degree distributions for conventional empirical networks~\cite{Masuda2018PhysRevE}. The results are qualitatively similar when one calculates the weighted degree of the $i$th node as $\sum_{j=1; j\neq i}^N |C_{ij}|$ or  $\sum_{j=1; j\neq i; C_{ij} > 0}^N C_{ij}$. Therefore, heterogeneous degree distributions of the correlation network are not an artifact of the thresholding or other operations for creating networks from correlation data, at least to some extent.

\subsection{Community detection\label{sec:community}}

Community structure in networks is a partition of the nodes into (generally non-overlapping) groups that are internally well connected and sparsely connected across.
One can detect communities with many different algorithms, and one popular family of methods, despite some shortcomings \cite{Peel2022NatComm}, is modularity maximization \cite{Porter2009AmerMathSoc,Fortunato2010PhysRep}, which aims at placing  a higher-than-expected number of edges connecting nodes within the same community. Modularity with a resolution parameter $\gamma$ is defined by
\begin{equation}
Q = \frac{1}{N\overline{k}} \sum_{i, j=1}^N
\left(A_{ij} - \gamma\frac{k_i k_j}{N\overline{k}}\right)
\delta_{g_i, g_j},
\label{eq:Q}
\end{equation}
where we remind that $A_{ij}$ is the entry of the empirical adjacency matrix, $g_i$ is the community in which the $i$th node is placed by the current partition, and $\delta$ is again the Kronecker delta.
Note the presence of the term $k_i k_j / N\overline{k}$ coming from Eq.~\eqref{eq:cm}, which signifies the use of the configuration model as standard null model in the ordinary modularity for networks. Approximate maximization of $Q$ by varying $g_1$, $\ldots$, $g_N$ for a given network identifies its community structure.

Community detection, in particular modularity maximization, is desirable for correlation network data, too. Given that the original correlation matrix has both positive and negative entries in general, a possible variant of modularity maximization for correlation networks is to maximize a modularity designed for signed networks. The modularity for signed networks may be defined as a weighted difference of the modularity calculated for the positive network (i.e., the weighted network only containing positively weighted edges) and the modularity calculated for the negative network (i.e., the weighted network only containing negatively weighted edges with the edge weight being the absolute value of the correlation)~\cite{Rubinov2011Neuroimage}. However, this procedure assumes that, in the null model, edges can be thought as independent (as in the model described by Eq.~\eqref{eq:cm}) and that positive edges can be randomized independently of negative edges. We have seen that both assumptions are clearly not valid for correlation matrices.

One can bypass the analogy with networks by directly computing and maximizing a modularity function that is appropriately defined for correlation matrices \cite{Macmahon2015PhysRevX}. A viable redefinition of modularity for correlation matrices is given by
\begin{equation}
Q^{\text{cor}} = \frac{1}{\mathcal{N}} \sum_{i, j=1}^N \left( \rho^{\text{sam}}_{ij} - \langle \rho_{ij} \rangle \right)
\delta_{g_i, g_j},
\label{eq:Q-correlation-matrix}
\end{equation}
where $\mathcal{N} = \sum_{i, j=1}^N \rho^{\text{sam}}_{ij}$ is a normalization constant, which is inessential to the maximization problem as long as it is positive. Matrix $\langle \rho\rangle$ should be a proper null model for the correlation matrix. Approximate maximization of $Q^{\text{cor}}$ provides optimal communities in correlation matrices. 
The crucial ingredient is the choice of the null model, $\langle \rho_{ij}\rangle$. Depending on what features of the original data one desires to preserve, the use of any of the models described in section~\ref{sub:null} is in principle legitimate, e.g., the white-noise (identity matrix) model $\rho^{\text{MG1}}_{ij}=\delta_{ij}$, the noise-only  model $\rho^{\text{MG2}}_{ij}$, the noise+global model $\rho^{\text{MG3}}_{ij}$, or the correlation matrix configuration model $C^{\text{cm}}_{ij}$. However, note that in the summation in Eq.~\eqref{eq:Q-correlation-matrix} some terms must be positive and some must be negative, but at the same time not dominated by noise, in order to identify a nontrivial community structure, i.e., one different from a single community enclosing all nodes. For example, if all the entries of the empirical correlation matrix, $\rho^{\text{sam}}$, are positive, then the null model $\rho^{\text{MG1}}$ will keep all terms in the summation in Eq.~\eqref{eq:Q-correlation-matrix} non-negative, and the resulting optimal partition will be a single community~\cite{Macmahon2015PhysRevX}.
By contrast, the use of $\rho^{\text{MG2}}$ removes the noisy component of the sample correlation matrix and allows to detect noise-filtered communities, unless a global trend is present. If a global trend is present, then all the entries of the filtered matrix in Eq.~\eqref{eq:PCA} are positive, preventing communities from being detected. When all the terms in the summation in Eq.~\eqref{eq:Q-correlation-matrix} are non-negative, the resulting optimal partition is a single community~\cite{Macmahon2015PhysRevX}, incidentally showing the limitation of PCA for the community detection task. In presence of such a global trend, one obtains the best results by using $\rho^{\text{MG3}}$, which uncovers group-specific correlations~\cite{Macmahon2015PhysRevX,almog2015mesoscopic,anagnostou2021uncovering,zema2021mesoscopic}. 
Having maximized the modularity
%
%
guarantees that the identified community structure is `optimally contrasted’, with necessarily positive overall residual correlations (with respect to the null model) inside each community and necessarily negative overall residual correlations across different communities. 
Modularity maximization using $\rho^{\text{MG3}}$ has successfully revealed nontrivial community structure in time series of financial stock prices~\cite{Macmahon2015PhysRevX,almog2015mesoscopic,zema2021mesoscopic}, credit default swaps~\cite{anagnostou2021uncovering}, single-cell gene expression profiles~\cite{mircea2022phiclust}, and neuronal activity~\cite{almog2019uncovering}. The last example is expanded below.

\emph{Application example:} Almog et al.~\cite{almog2019uncovering} applied RMT-based community detection, defined via the maximization of the modularity given by Eq.~\eqref{eq:Q-correlation-matrix}, to the empirical correlation matrix obtained from single-neuron time series of gene expression in the biological clock of mice. The recording was made from the suprachiasmatic nucleus (SCN), located in the hypothalamus. The biological clock is a highly synchronized brain region that is yet adaptable, for example, to external light stimuli and their seasonal variations. Therefore, the research focus was on the identification of both positive (excitatory, phase-coherent) and negative (inhibitory, phase-opposing) interactions among constituent neurons. They showed that methods based on dichotomization using a global threshold, as well as `naive’ community detection methods using the ordinary network-based modularity (i.e., Eq.~\eqref{eq:Q}), fail to identify groups of neurons that are internally positively correlated, and negatively correlated across. On the other hand, the maximization of the RMT-based modularity (i.e., Eq.~\eqref{eq:Q-correlation-matrix}), with the null model given by $\langle\rho\rangle= \rho^{\rm MG3}$ in Eq.~\eqref{eq:MG3}, successfully found community structure by filtering out both the neuron-specific noise and the system-wide dependencies that obfuscate the presence of underlying modules in the SCN. 
Their study uncovered two otherwise undetectable, negatively correlated populations of neurons (specifically, a spatially inner population and an outer one, both with left-right symmetry), whose relative size and mutual interaction strength were found to depend on the photoperiod~\cite{almog2019uncovering}. In particular, the average residual intra-community correlation was significantly higher in short photoperiods (e.g., winter) than in long photoperiods (e.g., summer). In contrast, the residual inter-community correlation was lower in short photoperiods than in long photoperiods.
A MATLAB package for the calculation of the null models used in the paper is available~\cite{Mel-Matlab-Rmt}.

\subsection{Clustering coefficient\label{sec:clustering}}

Clustering coefficients measure the abundance of triangles in a network~\cite{Caldarelli2007book, Barabasi2016book, Newman2018book, Menczer2020book}. A dominant definition of clustering coefficient for unweighted networks, denoted by $\tilde{C}$, is given by~\cite{Watts1998Nature}
\begin{equation}
\tilde{C} = \frac{1}{N} \sum_{i=1}^N \tilde{C}_i,
\end{equation}
where $\tilde{C}_i$ is the local clustering coefficient at the $i$th node given by
\begin{equation}
\tilde{C}_i = \frac{(\text{number of triangles involving the } i \text{th node})}
{k_i(k_i-1)/2}.
\label{eq:C_i}
\end{equation}
The denominator of the right-hand side of Eq.~\eqref{eq:C_i} is a normalization constant to ensure that $0\le \tilde{C}_i \le 1$. 

One often measures clustering coefficients for both unweighted and weighted correlation networks. There are various definitions of weighted clustering coefficients~\cite{Saramaki2007PhysRevE, WangGhumare2017NeuralComput}. One definition~\cite{ZhangHorvath2005StatApplGenetMolBiol} is given by $\tilde{C}^{\rm wei,Z} = N^{-1} \sum_{i=1}^N
\tilde{C}_i^{\rm wei,Z}$, where 
\begin{equation}
\tilde{C}_i^{\rm wei,Z} = \frac{1}{\max_{i^{\prime}j^{\prime}} w_{i^{\prime}j^{\prime}}}
\frac{\sum_{\substack{1\le j,\ell\le N\\ j,\ell\neq i}} w_{ij}w_{i\ell} w_{j\ell}}
{\sum_{\substack{1\le j,\ell\le N\\ j,\ell\neq i; j\neq \ell}} w_{ij}w_{i\ell}},
\label{eq:C_i^Zhang}
\end{equation}
and $w_{ij}$ $(= w_{ji} \ge 0)$ is the weight of edge $(i, j)$.

Many empirical networks show large unweighted or weighted clustering coefficient values, and correlation networks are no exception. However, as we pointed out in section~\ref{sub:dichotomizing}, a high clustering coefficient of the correlation network is at least partly due to pseudo correlation.

Given this background, clustering coefficients for correlation matrices were proposed using a similar idea to the case of modularity directly defined for correlation matrices~\cite{Masuda2018FrontNeuroinfo}.
Because correlation matrices are naturally clustered if we dichotomize on the Pearson correlation matrix, the authors used the three-way partial correlation coefficient or partial mutual information to partial out the effect of a common neighbor of nodes $j$ and $\ell$ (say $i$) to quantify partial connection between $j$ and $\ell$. In other words, we measure the connectivity between neighbors of $i$ by, for example, the partial correlation coefficient $\rho(X_{j}, X_{\ell} \mid X_i)$, which we abbreviate as $\rho_{j \ell \mid i}$; the partial correlation coefficient is defined by Eq.~\eqref{eq:three-way-partial-corr}.
Because there is no clear notion of neighborhood for correlation matrices, we need to consider all triplets of different nodes, ($i$, $j$, $\ell$). Then, as for the definition of the original clustering coefficient for networks, they took the average of $\rho_{j \ell \mid i}$ over the $i$th node's neighbors $j$ and $\ell$ to define a local clustering coefficient for $i$. For example, we define a local clustering coefficient for node $i$ as a weighted average by
\begin{equation}
C_i^{\rm cor,A} = \frac{\sum_{\substack{1\le j < \ell \le N\\ j, \ell \neq i}} \left|\rho_{i j} \rho_{i \ell} \rho_{j \ell \mid i}\right|}
{\sum_{\substack{1\le j < \ell \le N\\ j, \ell \neq i}} \left|\rho_{i j} \rho_{i \ell} \right|}.
\label{eq:C_i^A}
\end{equation}
Finally, as in the case of the clustering coefficient for networks, we define the global clustering coefficient by $C^{\rm cor, A} = \sum_{i=1}^N C_i^{\rm cor, A} / N$. This method borrows the idea of clustering coefficient from complex network studies and tailors it for correlation matrix data. Clustering coefficients $C_i^{\rm cor, A}$ and $C^{\rm cor, A}$ already partial out the effect of pseudo correlation between $X_j$ and $X_{\ell}$ due to $X_i$. However, we can still compare the observed clustering coefficient values against those for null models to validate whether or not the clustering coefficient values for the given data are significantly different from those for the null model~\cite{Masuda2018PhysRevE}.

\textit{Application example}: Masuda et al.\ searched for possible association of $C_i^{\rm cor,A}$, $C^{\rm cor,A}$, and similar clustering coefficients for correlation matrices with the age of human participants in fMRI experiments~\cite{Masuda2018FrontNeuroinfo}. They used publicly available resting-state fMRI data from the brains of healthy adults with a wide range of ages. The nodes were defined by  a commonly used brain atlas consisting of $N=30$ regions of interest. They found that the global clustering coefficients, such as $C^{\rm cor,A}$, declined with age. The correlation between the age and $C^{\rm cor,A}$ (and a variant of $C^{\rm cor,A}$) was stronger than that between the age and conventional clustering coefficients for general unweighted and weighted networks combined with both Pearson and partial correlation networks. Furthermore, the proposed local clustering coefficients were more strongly negatively correlated with age than the conventional clustering coefficients for general networks.

\section{Software\label{sec:software}}

In this section, we introduce freely available code useful for analyzing correlation networks. Obviously, one can apply software for analyzing general unweighted or weighted networks after thresholding (and optionally further dichotomizing) the given correlation matrix data. There are numerous packages for unweighted and weighted network analysis, which we do not mention here without a few exceptions.

In gene correlation network analysis, Weighted Correlation Network Analysis (WGCNA) is a popular freely available R software package \cite{Langfelder2008Bioinfo}. WGCNA provides various outputs, such as community structure, weighted clustering coefficients, and visualization. WGCNA transforms the original edge weight, denoted by $w_{ij}$ for edge $(i, j)$, by a so-called soft thresholding transformation, i.e., by $\left| w_{ij} \right|^{\beta}$, where $\beta \ge 1$ is a parameter\footnote{The soft thresholding here, coined in~\cite{Langfelder2008Bioinfo}, is different from the same term defined in section~\ref{sub:weighted}. It also does not belong to thresholding in the sense used in this article (defined in section~\ref{sub:dichotomizing}).}, such that one obtains an unsigned weighted network. Phiclust~\cite{mircea2022phiclust} is another community-detection tool for correlations from single-cell gene expression data, derived from RMT (i.e., Wishart ensemble as described in section~\ref{sub:null}). It can be used to identify cell clusters with non-random substructure, possibly leading to the discovery of previously overlooked phenotypes. Phiclust is written in R and is freely available at Github \cite{Phiclust-Github} and 
Zenodo under GNU General Public License V3.0 \cite{Phiclust-Zenodo}.

The Brain Connectivity Toolbox is a MATLAB toolbox for analyzing networks~\cite{Rubinov2010Neuroimage}. It is also implemented in Python. Despite the name, many of the functions provided by the Brain Connectivity Toolbox can be applied to general network data, and many people outside neuroscience also use it. In relation to correlation networks, this toolbox is particularly good at handling weighted and signed networks, such as their degree, clustering coefficients, and community structure.

The graph-tool module in Python provides powerful network analysis and visualization functionality~\cite{peixoto_graph-tool_2014}. In relation to correlation networks, graph-tool is particularly strong at network inference based on stochastic block models.

The bootnet package in R can be used for estimating psychological networks using graphical lasso, with a unique feature of being able to assess the accuracy of the estimated network \cite{Epskamp2018BehavRes}. This package can be used for other types of correlation matrix data as well. Also see \cite{Kuismin2017WileyInterdisciplinaryRevComputStat,Epskamp2018BehavRes} for various R packages for sparse and related estimation of the precision and covariance matrices. For example, the qgraph package can also generate networks using the graphical lasso and visualize Pearson and partial correlation matrices~\cite{Epskamp2012JStatSoftware}, with beginner-friendly tutorials (e.g., \cite{Gabriel2021blog}).

Graphical lasso is also implemented in Python, through the GraphicalLassoCV function in the scikit-learn package~\cite{scikit-learn,scikit-learn-code}. The sklearn.covariance package also contains other functions for covariance estimation such as covariance shrinkage. Table 2 of \cite{Borsboom2021NatRevMethodsPrimers} lists other code packages for estimating graphical models as well as different models. 

The Covariance Estimators package in Python, developed by Lucibello and Ib\`a\~nez-Berganza~\cite{Lucibello}, contains several utilities for denoising empirical covariance matrices. In particular, it implements various variants of PCA, linear shrinkage, graphical lasso, FA, early-stopping gradient ascent algorithms, and the RMT-based optimal RIE proposed in~\cite{ibanez2023noise} (see section~\ref{sec:RMT}).

Papers \cite{Macmahon2015PhysRevX,Vasa2022NatRevNeurosci} contain references to Python, MATLAB, and R codes for generating null models of  correlation matrices discussed in section~\ref{sub:null}. For example, the ``spatiotemporal modeling tools for Python'' contains functions to generate null model correlation matrices such as the H--Q--S model (named Zalesky matching model in their package) and methods through generating surrogate time series~\cite{Shinn2023NatNeurosci} (see section~\ref{sec:homogeneous}). Another package in the list is the Scola, in Python, which generates the H--Q--S model and the correlation matrix configuration model~\cite{Kojaku2019ProcRSocA} (see section~\ref{sec:configuration}).
Finally, a MATLAB package by MacMahon~\cite{Mel-Matlab-Rmt} implements the null models based on RMT, $\rho^{\rm MG2}$, and $\rho^{\rm MG3}$, derived in~\cite{Macmahon2015PhysRevX} from the Wishart ensemble (see section~\ref{sec:RMT}).

\section{Outlook}

\subsection{Recommended practices\label{sub:recommendations}}

We have reviewed various techniques to obtain and analyze networks generated from correlation matrix data, which naturally arise across many domains. Sections~\ref{sec:appl} and~\ref{sec:math} emphasize for readers that there is not a single dominant method.
We also highlighted good practices and pitfalls of individual methods. Na\"ive applications of network generation and analysis methods to correlation matrix data can easily yield flawed results. We should be careful about both how to generate correlation networks and how to analyze them. We recommend the following practices.

First, in resonance with previous reports, we explained that a simplistic dichotomizing, which is widely used, is problematic for multiple reasons (see section~\ref{sub:dichotomizing}). Therefore, if you threshold your correlation matrices to create networks, which may or may not be followed by dichotomization, do either of the following: (i) report your downstream network analysis results across a wide range of the threshold value; (ii) use a method designed to investigate a range of thresholds altogether (e.g., persistent homology); (iii) devise and use a network index that is robust with respect to the choice of the threshold value; and/or (iv) still use the simplistic thresholding but combine it with a proper null model. We showed an example of (i) at the end of section~\ref{sub:dichotomizing}. All of these practices are heuristic, but they are substantially better than simply using a single threshold value, reporting network analysis results, and skipping the examination of robustness.

That said, we do not have much knowledge about item (iii) regarding which network indices one should use. For example, the values of many network indices probably depend on the threshold value, which directly affects the number of edges \cite{Zalesky2012Neuroimage}. However, in most cases, we are interested in comparing network analysis results between different cases, such as between disease and control groups, between empirical and randomized data, or between individuals of different ages. Then, the group difference or ranking among individuals may be robust enough when one varies the threshold value. Investigating robustness of correlation network analysis outcomes with respect to the threshold warrants more work.

To illustrate item (iv), we recall that correlation networks have high clustering no matter what data we use. However, a proper null model (e.g., shuffling of $\{ x_{i\ell} \}$) will also produce dichotomized correlation networks with high clustering. Therefore, by comparing the results with those for the null model, one can avoid wrong conclusions such as that almost all empirical correlation networks have high clustering. We recall that null models for networks, most famously the configuration model, are not a proper null model for correlation networks because they generally do not originate from correlation matrices and do not generally match key properties of typical correlation matrices. See section~\ref{sub:null} for proper choices. 

Our second, alternative recommendation is to resort to other methods, such as weighted correlation networks, graphical lasso (which is a partial correlation network method) and its variants, and covariance shrinkage, which avoid thresholding. Nonetheless, these methods usually require some arbitrary decisions by the users, such as setting hyperparameter values. Therefore, assessing robustness with respect to such choices remains important. For example, with weighted correlation networks without thresholding, one usually chooses between whether to force negative correlation values to zero, to keep them by taking the absolute value of the correlation, or to treat them as signed networks. Few papers have investigated different cases to check robustness of the subsequent network analysis results. Furthermore, because these different operations have been main options for a long time, it may be beneficial to pursue quantification of weighted networks that would provide results that are robust with respect to this methodological choice.
 
Our third recommendation is to avoid transforming correlation matrix data into networks but yet carry out downstream analysis analogous to network analysis. However, we recommend doing so only for properties whose definition does not make (implicit) assumptions that are violated by correlation matrices, such as the assumption of independent matrix entries that the ordinary modularity function in Eq.~\eqref{eq:Q} implicitly makes. In presence of such unverified assumptions, we recommend either appropriately revising the definition of the property, e.g.\ as in the modified modularity in Eq.~\eqref{eq:Q-correlation-matrix}, or dismissing the property altogether.
%
%
In section~\ref{sub:inspired}, we showcased some such methods including two example analyses. This type of analysis is available for at least the degree, modularity maximization, and clustering coefficients. Then, one can evade thresholding or thoroughly examining various threshold values. Future work should generalize this approach to other structural properties and analysis methods formulated for networks. Examples include various node centrality measures, motifs, community detection methods apart from modularity maximization, rich clubs, fractals, and simplicial complexes. In many cases, the configuration model is a standard choice of null model when constructing a network algorithm, such as a community detection algorithm. However, while we now have a reasonably long list of null models for correlation matrices (see section~\ref{sub:null}), it is not known whether the configuration model for covariance matrices or a different null model described in  section~\ref{sub:null} should be a standard choice when constructing an algorithm for correlation matrices inspired by network analysis. Answering this question needs systematic comparison studies of downstream analyses across various null models for correlation matrices.

\subsection{Directions for future research}

In this section, we identify some promising directions for future research.

Reproducibility of correlation networks arising from common approaches is a major practical issue, as has been pointed out in the literature in psychology and psychotherapy (e.g.~\cite{Fried2017PersPsycholSci,Forbes2017JAbnormalPsychol}) and microbiome analysis~\cite{Goberna2022SoilBiolBiochem}. In these research fields and others, it is often the case that only relatively few samples are available given the size of matrix and network, $N$, we wish to analyze.
Especially if the number of samples, $L$, is smaller than or close to $N$, the partial correlation matrix and spectra of random matrices carry large variation, in part because they are inherently rank deficient. From the statistical inference point of view, it is not sound to try to infer many parameters, such as entries of the covariance matrix, from relatively few observations. The recognition of such phenomena led to the idea of covariance selection and various other methods. The amount of data needed for reliably estimating correlation networks, i.e., power analysis in statistics, should be further pursued for various correlation matrix data \cite{Epskamp2018BehavRes}. 
The development of methods to help practitioners use correlation networks better (e.g., by providing uncertainty quantification or clarifying the various noise trade-offs) can be transformational. Despite these challenges, there is a pressing need to understand complex systems of a very high dimension (i.e., $N \gg L$) with correlational data. One approach to this problem is to formulate estimation of large correlation networks as a computational rather than statistical challenge, as a problem to be solved under runtime and memory constraints, and to search feasible solutions in combination with machine learning~\cite{Becker2023NatComputSci}.
How to reconsile the statistical and computational types of approach and deepen usage of machine learning and artificial intelligence techniques to correlation network analysis may be a beneficial research direction.

A key outstanding question is the treatment of low-correlation edges. On one hand, we have surveyed attempts to remove ``noise'' edges (for example by thresholding or graphical lasso), which is supposed to improve the overall signal-to-noise ratio of the graph representation. Sparser models are more parsimonious, easier to process quickly and with a lower memory footprint, and amenable to a range of network science analysis tools. On the other hand, one can argue that getting rid of low-correlation edges risks losing valuable information (see section~\ref{sub:dichotomizing}). In fact, it has been shown in the neuroscience context that the low-correlation edges alone can have substantial predictive power~\cite{Bassett2012Neuroimage,cole2012global, santarnecchi2014efficiency}.

A related question is how to use \emph{a priori} domain knowledge to choose appropriate preprocessing steps, such as what threshold to apply and whether or not to dichotomize.
For example, dichotomizing may be more appropriate when the \emph{a priori} belief is that nodes are either coordinating or not, with no appreciable difference in the degree of coordination when two nodes coordinate.
As another example, one could use RMT on domain-relevant distributions to compare the dominant eigenvectors or eigenvalues before and after thresholding. This exercise may provide guidance on when thresholding is unlikely to have adverse effects. 

Another viable alternative to the current focus on trying to recover and analyze the most accurate possible correlation matrix may be to treat the constructed correlation network as inherently uncertain and to regard it as a sample from a distribution of possible matrices as part of the analysis. For example, when assessing community structure, it may make sense to focus on structures that appear consistently across samples of correlation networks drawn from the estimated distribution, even when exact correlation values (perhaps especially the weaker correlations) considerably vary from sample to sample. Although there are established ways to do this for general networks~\cite{newman2020bayesian}, modeling of correlation networks and their substructures by their probability distributions is still a new idea \cite{Raimondo2021PhysRevE} and needs further development. Such approaches may leverage existing work on null models for correlation networks, for example, as priors when forming a posterior distribution to sample from. On the other hand, some studies have documented the stability of the detected correlation-induced communities across time and their robustness under change of temporal resolution~\cite{Macmahon2015PhysRevX,almog2015mesoscopic,almog2019uncovering,anagnostou2021uncovering,zema2021mesoscopic}.


There are many multilayer network data sets, including multilayer correlation matrix data sets, and various data-analysis methods for them~\cite{Boccaletti2014PhysRep,Kivela2014JCompNetw,Bazzi2016MultModelSimul,Bianconi2018book, Artime2022Cambridge, Dedomenico2023NatPhys}. Examples include brain activity, where different layers of correlation matrices correspond to, for example, 
different frequency bands \cite{Brookes2016Neuroimage,Tewarie2016Neuroimage,Buldu2017NetwNeurosci,Dedomenico2016FrontNeurosci}, or brain activity during different tasks \cite{Dedomenico2017GigaSci}. In gene co-expression networks, different layers correspond to, for example, different levels of co-expression between gene pairs \cite{DorantesGilardi2020ApplNetwSci} or different tissue types such as different organs
\cite{Russell2023PlosComputBiol}. Overlaying different methods to construct correlation networks from one data set in each layer is another method to construct multilayer correlation matrices~\cite{Musmeci2017Complexity}. There are methods for analyzing multilayer correlation matrices and networks such as multilayer community detection algorithms \cite{Bazzi2016MultModelSimul,Russell2023PlosComputBiol}.
However, methods that exploit the mathematical nature of correlation matrix data are still scarce and left for future work. Furthermore, multilayer networks are a major representation of temporal network data, where each layer corresponds to one time window \cite{Mucha2010Science,Bazzi2016MultModelSimul,Masuda2020book}. Therefore, methods for analyzing multilayer correlation network data will also contribute to analysis of time-varying correlation network data.

Similar to other studies, we emphasize the potentially negative effects of thresholding, motivating our explanation of other methods for constructing correlation networks. However, thresholding also has positive effects such as reducing false positives by discarding edges with small weights including the case of correlation networks. Such positive effects of thresholding may be manifested in multilayer data. For example, aggregating layers in a multilayer network and dichotomizing the aggregated edge weight can enhance detectability of multilayer communities compared to no dichotomizing under certain conditions~\cite{TaylorShai2016PhysRevLett,TaylorCaceres2017PhysRevX}. Furthermore, some layers in a multilayer network may be more informative than other layers. While these arguments are for general multilayer networks, many of them may directly apply to multilayer correlation matrices.

For a given $N$, the set of covariance matrices constitutes the positive semidefinite cone, which is convex. Similarly,
the set of full-rank correlation matrices, which is a strict subset of full-rank covariance matrices, is called the elliptope \cite{Tropp2018DiscreteComputGeom,Thanwerdas2021Lncs}. Positive semidefinite cones and elliptopes are manifolds and have their own geometric structure, which have been suggested to be useful for measuring the similarity between pairs of covariance or correlation matrices. Quantitative comparison of two covariance and correlation matrices is useful for various tasks such as fingerprinting of individuals, anomaly detection, and change-point detection in multivariate time series data. A straightforward way to measure the distance between two covariance/correlation matrices is to use a common matrix norm such as the Frobenius norm (i.e., $\sqrt{\sum_{i=1}^N \sum_{j=1}^N \left| \rho^{(1)}_{ij} - \rho^{(2)}_{ij} \right|^2}$ in the case of correlation matrices, where $\rho^{(1)}$ and $\rho^{(2)}$ are two correlation matrices). However, research (in particular in neuroimaging studies) has suggested that geodesic distance measures respecting the structure of these manifolds is better at, for example, fingerprinting of individuals from fMRI data \cite{Venkatesh2020Neuroimage,YouPark2021Neuroimage,Abbas2021BrainConn}. In these geodesic distance measures, one considers the tangent space at a given point $x$ on the manifold, which corresponds to a correlation/covariance matrix. The so-called exponential map provides a one-to-one mapping from the straight line segment on the tangent space, which is essentially the Euclidean space, to the geodesic from $x$ to $y$ on the manifold. The logarithm map is the inverse of the exponential map. The geodesic distance between $x$ and $y$ is the length of the geodesic and has a practical matrix algebraic formula. Multiple reasonable definitions of such geodesic distances exist \cite{YouPark2021Neuroimage}. See \cite{Pennec2006IntJComputVis,Rahim2019MedImageAnal,YouPark2021Neuroimage} for mathematical expositions. Although these techniques are not for correlation networks but for matrices, they may potentially benefit understanding and algorithms for correlation networks. For example, can we understand null models of correlation matrices as a projection onto a submanifold of the entire elliptope? What are geometric meanings of thresholding, dichotomizing, and other operations to create correlation networks? Do we benefit by measuring distances between correlation networks rather than between correlation matrices? 

We mentioned examples of microbiome and bibliometric co-occurrence networks as variants of correlation networks in sections~\ref{sub:microbiome} and \ref{sub:bibliometric}, respectively. For example, let $x_{i\ell} = 1$ if the $i$th researcher is an author of the $\ell$th paper in the database and $x_{i\ell} = 0$ otherwise. Then, the number of papers that the $i$th and $j$th researchers have coauthored, which are co-occurrence events, is equal to
$\sum_{\ell=1}^L x_{i\ell} x_{j\ell}$, where $L$ is the number of papers in the entire data. This quantity is a non-standardized covariance, which one can analyze as a correlation network. In fact, the original data, i.e., $N\times L$ matrix $(x_{i\ell})$, is a bipartite network, in which one part consists of $N$ nodes representing the researchers, and the other part consists of $L$ nodes representing the papers. An edge exists between an author node and a paper node if and only if $x_{i\ell} = 1$. Then, we can interpret the $N \times N$ co-occurrence, or correlation, network whose $(i, j)$ entry is given by $\sum_{\ell=1}^L x_{i\ell} x_{j\ell}$ as a one-mode projection of the bipartite network. This viewpoint provides research opportunities. It is known that one-mode projection introduces various biases \cite{Latapy2008SocNetw}. For example, one-mode projection inflates the clustering coefficient \cite{Newman2001PhysRevE-collabo1, Guillaume2004InfoProcLett, Ramasco2004PhysRevE}, which is in fact consistent with the finding that correlation networks even from random data would have a high clustering coefficient (section~\ref{sub:dichotomizing}). One strategy to avoid such biases is to analyze the data as a bipartite network \cite{Latapy2008SocNetw}. Therefore, bipartite network analysis may be equally useful for understanding correlation structure of continuous-valued data, i.e., an $N\times L$ matrix $(x_{i\ell})$, $x_{i\ell} \in \mathbb{R}$, which we have mostly dealt with in the present article. Establishing mapping of the continuous-valued data to a bipartite network is a first natural step toward this goal. A framework that both handles bipartite signed networks and provides a statistically validated one-mode projection \cite{Gallo2025arxiv} may be fruitfully extended to this direction.

We have confined the present review to pairwise correlation. It is possible to quantify co-fluctuations among more than two nodes and analyze their system-wide organization by hypergraph \cite{Gallo2024NatCommun} or simplicial complex \cite{Santoro2023NatPhys, Santoro2024NatCommun} techniques. In fact, pursuit of higher-order correlation among three or more variables is itself not new; various information-theoretic and statistical methods have been available to quantify higher-order correlation in data for more than two decades, particularly in computational neuroscience \cite{Amari2003NeuralComput, Schneidman2003PhysRevLett, YuYang2011JNeurosci, Shimazaki2012PlosComputBiol, Varley2023CommunBiol, Varley2023PNAS}. As a network represents a set of pairwise correlations, a hypergraph or simplicial complex \cite{Battiston2020PhysRep, Battiston2021NatPhys, Bianconi2021book} (also see section~\ref{sub:persistent-homology} for references for the latter) represents a set of higher-order correlations. Further developing methods for analyzing data with higher-order correlation using hypergraph and simplicial complex techniques would be fruitful.

One complication that has not received enough attention is that many in-practice comparisons involve ensembles of observed networks rather than single networks. This is the case in most fMRI studies where networks are used. When working with an ensemble of networks, one must make various decisions, such as whether or not to ensure that edge density is constant across networks (see section~\ref{sub:dichotomizing} for the absolute versus proportional threshold). The development of mathematical theories for how to construct correlation-based networks for ensembles may be helpful because most null models and other tools are only oriented toward single networks. Multilayer approach and geometric approaches to correlation networks and matrices, both of which cater to between-network/matrix comparisons, are promising paths towards this goal.

A graphon is a symmetric measurable function $W: [0, 1]^2 \to [0, 1]$. Given $W$, we generate dense graphs in which there is an edge between the $i$th and $j$th nodes with probability $W(u_i, u_j)$, where each $u_i$ $\forall i$ independently obeys the uniform density on $[0, 1]$ \cite{Lovasz2012book}. In network science, assigning a node weight, either from a probability distribution or empirical data, and connecting two nodes probabilistically as a function of the two node weights has been a major method to generate networks \cite{Bianconi2001PhysRevLett,Bianconi2001EurophysLett,Goh2001PhysRevLett,Caldarelli2002PhysRevLett,Boguna2003PhysRevE,Masuda2004PhysRevE-threshold,Perra2012SciRep}. Basic correlation networks are equivalent to an extension of this class of network models where $u_i$ is an $L$-dimensional vector of the $i$th feature from $L$ samples, and $W$ is a criterion with which to determine the edge. In fact, similar to the construction of correlation networks by dichotomizing, dichotomizing functions have been used as $W$ to generate networks with power-law degree distributions from scalar node weights that do not have to obey long-tailed distributions \cite{Caldarelli2002PhysRevLett,Masuda2004PhysRevE-threshold,Masuda2005PhysRevE,Masuda2006SocNetw,Cantwell2020PhysRevE}.
%
%
Importing mathematical frameworks and methods from graphon-related models, such as the limit of a sequence of dense networks, to correlation network analysis may be an interesting idea. Those frameworks may be able to provide null models for correlation networks or give more mathematical foundations of correlation networks.

\subsection{Final words}

We have seen that there are various research fields in which they collect and analyze correlation networks. We have also seen that some particular analysis techniques are heavily studied in one field, but others are preferred in different fields. For example, random matrices for sample correlation matrices~\cite{livan2018introduction,potters2020first} have particularly been used in financial data studies, including econophysics~\cite{Bun2017PhysRep,Laloux1999PhysRevLett,Plerou1999PhysRevLett,Macmahon2015PhysRevX,almog2015mesoscopic,anagnostou2021uncovering,zema2021mesoscopic}, while they have been applied much less, and only more recently, in neuroscience~\cite{almog2019uncovering,ibanez2023noise}. As another example, a majority of studies of temporal correlation networks has been done in network neuroscience under the name of dynamic functional connectivity/networks. However, very often, methods for analyzing correlation networks developed in one research field do not rely on particularities of the field and are therefore transferable to other research fields. While such cross-fertilization has been ongoing and advocated \cite{Borsboom2021NatRevMethodsPrimers}, we emphasize that much more of it will be useful for furthering correlation network analysis and applications. By the same token, studies directly comparing objectives and performances of methods used in different research domains will be valuable.

Cross-fertilization is also desirable within theoretical fields. Statisticians and non-statisticians tend not to know each other's work and publish in different types of journals. Statisticians tend to start with research questions that are ideally asked and answered by statistical hypothesis testing or Bayesian methods. Therefore, they would develop methods for correlation networks with which the data analysis results can be statistically tested. In contrast, non-statistically-focused researchers including many applied mathematicians, physicists, and computer scientists tend to focus on network analysis techniques which may be heuristic or reflect analogies to other processes, many of which have been proven to be powerful in different settings. Because there are already many useful network analysis methods, if one can exploit them in correlation data analysis, these types of researchers, including the present authors, would expect that it is a beneficial connection. We tried to cover both types of approaches to correlation networks as much as possible in this article. We believe that more discussion between these different perspectives on correlation networks will drive further developments of both types of approach. See for example \cite{Porter2017arxiv} for related discussion.

\section*{Acknowledgements}

We thank Sarah Muldoon for discussion. N.M. acknowledges support from National Institute of General Medical Sciences (under grant no.\ R01GM148973),
%
%
the Japan Science and Technology Agency (JST) Moonshot R\&D (under grant no.\ JPMJMS2021), the National Science Foundation (under grant nos.\ 2052720 and 2204936), and JSPS KAKENHI (under grant nos.\ JP 21H04595 and 23H03414). 
P.J.M.\ and Z.M.B.\ acknowledge support from the Army Research Office (under MURI award W911NF-18-1-0244). P.J.M.\ also acknowledges support from the National Institute of Diabetes and Digestive and Kidney Diseases (under grant no.\ R01DK125860) and from the National Science Foundation (under grant no.\ 2140024). Z.M.B.\ also acknowledges support from the National Science Foundation (under grant no.\ 2137511).
D.G. acknowledges support by the European Union - NextGenerationEU - National Recovery and Resilience Plan (Piano Nazionale di Ripresa e Resilienza, PNRR), project `SoBigData.it - Strengthening the Italian RI for Social Mining and Big Data Analytics' - grant IR0000013 (n.\ 3264, 28/12/2021) (\url{https://pnrr.sobigdata.it/}), project `RECON-NET - Reconstruction, Resilience and Recovery of Socio-Economic Networks' - grant EP\_FAIR\_005, PE0000013 ``FAIR'' (PNRR M4C2 Investment 1.3), and project `MNESYS - A Multiscale integrated approach to the study of the nervous system in health and disease' - grant PE0000006 (DN. 1553 11.10.2022).
The content is solely the responsibility of the authors and does not necessarily represent the official views of any agency supporting this work.


\end{document}